\documentstyle[12pt,epsf]{article}
\begin{document}
%%%%%%%%%%%%%%%%%%%%%%%%%%%%%%%%%%%%%%%%%%%%%%%%%%%
\def\thefootnote{\fnsymbol{footnote}}
\begin{flushright}
KANAZAWA-97-13  \\ 
August, 1997
\end{flushright}
%\vspace{ .7cm}
\vspace*{2cm}
\begin{center}
{\LARGE\bf Neutralino decay in the $\mu$-problem solvable 
extra $U(1)$ models}\\
\vspace{1 cm}
{\Large Daijiro Suematsu}
\footnote[1]{e-mail: suematsu@hep.s.kanazawa-u.ac.jp}
\vspace {1cm}\\
{\it Department of Physics, Kanazawa University,\\
        Kanazawa 920-11, Japan}\\    
\end{center}
\vspace{1cm}
{\Large\bf Abstract}\\  
%%%%%%%%%%%%%%%%%Abstract%%%%%%%%%%%%%%%%%%%%%%%%%%%
We study the neutralino decay in the supersymmetric 
extra $U(1)$ models which can solve the $\mu$-problem. 
In these models the neutralino sector is extended at least into six
components by an extra $U(1)$ gaugino and a superpartner of a singlet 
Higgs. Focussing on its two lower mass eigenstates $\tilde\chi_2^0$
and $\tilde\chi_1^0$, decay processes such as
a tree level three body decay $\tilde\chi_2^0\rightarrow \tilde\chi_1^0
f\bar f$ and a one-loop radiative decay $\tilde\chi_2^0\rightarrow 
\tilde\chi_1^0\gamma$ are estimated. We investigate the 
condition under which the radiative decay becomes the dominant
mode and also numerically search for such parameter regions.
In this analysis we take account of the abelian gaugino kinetic term 
mixing. We suggest that the gaugino mass relation $M_W\sim M_Y$ may
not be necessary for the radiative decay dominance in the extra $U(1)$
models. 
\vspace{0.5cm}

\noindent
PACS number(s): 12.60.Jv, 14.80.Ly
\newpage
\setcounter{footnote}{0}
\def\thefootnote{\arabic{footnote}}
%%%%%%%%%%%%%%%%%%%%%%text%%%%%%%%%%%%%%%%%%%%%%
\section{Introduction}
Recently the standard model (SM) has been confirmed in the incredible accuracy
through the precise measurements at LEP.
Nevertheless, it has still not been considered as the fundamental theory of
particle physics and physics beyond the SM is eagerly explored.
Along this line the supersymmetrization of the SM is now considered
as the most promising extension \cite{n}.
However, even in this minimal supersymmetric standard model (MSSM) 
there remain some theoretically unsatisfactory features in addition
to the existence of too many parameters.
The famous one is known as the $\mu$-problem \cite{mu}. 
The MSSM has a supersymmetric Higgs mixing term $\mu H_1H_2$. 
To cause an appropriate radiative
symmetry breaking at the weak scale \cite{rad}, we should put $\mu\sim
O(G_F^{-1/2})$ by hand, where $G_F$ is a Fermi constant.
Although in the supersymmetric models its typical scale is generally 
characterized by the supersymmetry breaking scale $M_S$ which is 
usually taken as 1~TeV region, 
there is no reason why $\mu$ should be such a scale because it is 
usually considered to be irrelevant to the
supersymmetry breaking. The reasonable way to answer this issue is to
consider the origin of $\mu$-scale as some result of the
supersymmetry breaking \cite{musol}. 
One of such solutions is the introduction of a
singlet field $S$ and replace $\mu H_1H_2$ by a Yukawa type coupling
$\lambda SH_1H_2$. 
If $S$ gets a vacuum expectation value(VEV) of order
1~TeV as a result of renormalization effects on the soft supersymmetry 
breaking parameters, $\mu\sim O(G_F^{-1/2})$ will be realized dynamically 
as $\mu=\lambda\langle S\rangle$. 
As is well-known, such a scenario can be available by introducing a
$\kappa S^3$ term into the superpotential and
a lot of works have been done on this type of models \cite{singlet}, where 
the superpotential of $S$ is composed of the terms $\lambda SH_1H_2+\kappa
S^3$. At the price of the introduction of a new parameter $\kappa$,
a $\kappa S^3$ term can prohibit the appearance of a massless 
axion and also guarantee the stability of the potential for the scalar
component of $S$. The introduction of an extra $U(1)_X$ symmetry
which is broken by a SM singlet field $S$ can effectively play the 
same role of the introduction of the $\kappa S^3$ term \cite{sy}. 
A D-term for this $U(1)_X$ induces a quartic term of 
$S$ in the scalar potential. The axion is eaten by this extra
$U(1)_X$ gauge boson and disappears from the physical spectrum.
Moreover, this extra $U(1)_X$ automatically forbids the appearance 
of $\mu H_1H_2$ in the original Lagrangian
and also if we assume the unification of gauge coupling constants,
we need no new parameter like $\kappa$.
Thus the models  extended with an extra $U(1)_X$ symmetry can be
considered as one of the most simple and promising extensions of the MSSM.
Their phenomenological aspects have also been studied by various 
authors \cite{sy,grh,extra,string}.

The extra $U(1)_X$ models have an another interesting aspect
if they are supersymmetrized.
Their supersymmetrization introduces the extra neutralino 
candidates besides the ones of the MSSM, 
that is, an extra $U(1)_X$ gaugino $\lambda_X$ and a superpartner
$\tilde S$ of the singlet Higgs $S$.
The confirmation of the extra gauge structure is one of the main parts
of the study of extension of the SM.
It is well-known that the extra $U(1)_X$ gauge structure is often induced from 
more fundamental theory like superstring \cite{string}.
However, recent precise measurements at LEP and the direct search at Tevatron
suggest that the lower bound of the extra neutral
gauge boson is rather large and it may be difficult to find its
existence directly in near future \cite{exp}.
If the supersymmetry is the true story in nature, 
there may be a new possibility
to find its existence in the completely different way \cite{sue}.
Even if the mass of extra neutral gauge boson is too large to observe
its existence in the near future collider experiment, its superpartner sector 
may open the window to find its existence.
The study of neutralino sector is interesting
from the view point not only of the investigation of supersymmetry but also
of the search for the extra gauge structure.
In particular, we should note that the gauge coupling of this extra
$U(1)_X$ to ordinary matter fields is rather large compared with the ordinary
Yukawa couplings (instead of top Yukawa)\footnote{
It should also be noted that the Yukawa coupling $\lambda$ of $\lambda 
SH_1H_2$ can be large enough compared with the ordinary Yukawa couplings.}
and then the neutralino sector
can be substantially affected by this inclusion in  a suitable
parameter region. 

In this paper we treat the neutralino decay in the extra $U(1)_X$
models since it may be one of the important subjects along the above
mentioned direction.
The lightest neutralino is a candidate of the lightest supersymmetric
particle. Thus if R-parity is conserved, the neutralino decay modes such as 
$\tilde\chi_2^0\rightarrow\tilde\chi_1^0 f\bar f$ and
$\tilde\chi_2^0\rightarrow \tilde\chi_1^0\gamma$ are expected to 
appear as the subprocess of the decay of supersymmetric particles,
where $\tilde\chi_2^0$ and $\tilde\chi_1^0$ are the two lower
neutralino mass eigenstates. 
These decay processes have been calculated in the case of 
the MSSM under the suitable conditions \cite{mssmdecay}.

Recently, some attentions have been attracted to this process in relation 
to the CDF $ee\gamma\gamma+/\hspace{-2.7mm}E_T$ event \cite{cdf}.
Especially, related to this type of events, it seems to be 
a very interesting subject under
what condition $\tilde\chi_2^0\rightarrow \tilde\chi_1^0\gamma$
can become the dominant mode \cite{cdf,am}.
This is because it can give us the fruitful information on the
parameters of supersymmetric models as stressed in \cite{cdf}.
Since this type of process is a typical one which may be observed in near 
future, its detailed study in the $\mu$-problem solvable extra $U(1)_X$
models will be useful.
The estimation of the widths $\Gamma(\tilde\chi_2^0\rightarrow
\tilde\chi_1^0\gamma)$ and $\Gamma(\tilde\chi_2^0\rightarrow
\tilde\chi_1^0f\bar f)$ in the extra $U(1)_X$ models can be modified 
from that in the MSSM because there are new components 
$\lambda_X$ and $\tilde S$ contained in the neutralino mass eigenstate 
$\tilde\chi_i^0$.
Additionally, in the multi $U(1)$s models the abelian kinetic term mixing
can occur as suggested in refs. \cite{mixing,ms,mixing2}. 
As a result of this abelian kinetic term mixing, 
there is some changes in the interactions between neutralinos
and ordinary matter fields \cite{sue}.
This should be taken into account in the analysis of these processes.
Due to these effects the $\tilde\chi_2^0\rightarrow \tilde
\chi_1^0\gamma$ dominant condition is also expected to be altered from the
MSSM one.
If we take the lesson brought from the study of the CDF type 
event seriously, this analysis may give us an important information 
for the model building on the additional gauge structure and also 
the Planck scale physics.

The organization of this paper is the following.
In section 2, we present the examples of the $\mu$-problem solvable
extra $U(1)_X$ models derived from the superstring inspired $E_6$ models.
After that we give a brief review of the abelian gaugino mixing
whose effect is taken into account in the later analysis.
We also examine the neutral gauge boson and Higgs sector to constrain
the parameters of the models in terms of their present experimental mass 
bounds.
In section 3, mass eigenstates and their couplings to the matter fields of the
extended neutralino sector are studied. Based on these preparations
the decay widths $\Gamma(\tilde\chi_2^0\rightarrow\tilde\chi_1^0 f\bar f)$
and $\Gamma(\tilde\chi_2^0\rightarrow\tilde\chi_1^0\gamma)$ are
estimated. We also study under what condition the radiative decay mode
becomes the dominant one, which is crucially relevant to the CDF type event.
In section 4, these decay widths are numerically estimated
and we show what kind of parameter region is crucial for the radiative 
decay dominance. 
Section 5 is devoted to summary.

\section{Extra $U(1)_X$ models}
\subsection{$\mu$-problem solvable models}
There can be a lot of low energy extra $U(1)_X$ models.
In these models we are especially interested in $\mu$-problem solvable
extra $U(1)_X$ models. From such a point of view, it seems to be natural 
to examine the models which satisfy a condition mentioned in the 
introduction.
That is, the extra $U(1)_X$ symmetry should be broken by the VEV of the SM
singlet $S$ which has a coupling to the ordinary Higgs doublets $H_1$
and $H_2$ such as $\lambda SH_1H_2$.
In these models the $\mu$-scale is naturally related to the mass of the 
extra $U(1)_X$ boson and then they seem to be very interesting
from the phenomenological viewpoint too.\footnote{
There is also a possibility that the $\mu$-term is realized by a
nonrenormalizable term $\lambda (S\bar S/M_{\rm pl}^2)^nSH_1H_2$
because of some discrete symmetry \cite{discr}. In such a case 
$\langle S\rangle$ should be large in order to realize the appropriate 
$\mu$-scale. As a result there is not the low energy 
extra gauge symmetry which can be relevant to the present experimental 
front. Because of this reason we do not consider this possibility.} 
So we confine our attention to this class of models derived from the
superstring inspired $E_6$ models.

There are two classes of extra $U(1)_X$ models derived from superstring
inspired $E_6$ models.
The rank six models have two extra $U(1)$s besides the SM gauge structure.
They can be expressed as the appropriate linear combinations of
$U(1)_\psi$ and $U(1)_\chi$ whose charge assignments for ${\bf 27}$ of
$E_6$ are given in Table 1.
\begin{figure}[bt]
\begin{center}
\begin{tabular}{|c|c|c||c|c||c|}\hline
fields &$SU(3)\times SU(2)$& $Y$ & $Q_\psi$ & $Q_\chi$ & $Q_\eta$ \\
\hline
$Q$  &(3,2)&${1\over 3}$&$\sqrt{5\over 18}$&
$-{1\over\sqrt6}$&$-{2\over 3}$   \\    
$U^c$&$(3^\ast,1)$&$-{4\over 3}$&$\sqrt{5\over 18}$&$-{1\over\sqrt
6}$&$-{2\over 3}$ \\
$D^c$&$(3^\ast,1)$&${2\over 3}$&$\sqrt{5\over 18}$&${3\over\sqrt 6}$&
${1\over 3}$ \\
$L$  &(1,2)&$-1$&$\sqrt{5\over 18}$&${3\over\sqrt 6}$&${1\over 3}$ \\
$E^c$&(1,1)&2&$\sqrt{5\over 18}$&$-{1\over\sqrt 6}$&$-{2\over 3}$ \\
$H_1$&(1,2)&$-1$&$-2\sqrt{5\over 18}$&$-{2\over\sqrt 6}$&${1\over 3}$ \\
$H_2$&(1,2)&1&$-2\sqrt{5\over 18}$&${2\over\sqrt 6}$&${4\over 3}$ \\
$g$&(3,1)&$-{2\over 3}$&$-2\sqrt{5\over 18}$&${2\over\sqrt 6}$&${4\over 3}$ \\
$\bar g$&$(3^\ast,1)$&${2\over 3}$&$-2\sqrt{5\over 18}$&
$-{2\over\sqrt 6}$&${1\over 3}$ \\
$S$ &(1,1)&0&$4\sqrt{5\over 18}$&0&$-{5\over 3}$ \\
$N$ &(1,1)&0&$\sqrt{5\over 18}$&$-{5\over\sqrt 6}$&$-{5\over 3}$ \\\hline
\end{tabular}
\vspace*{.2cm}
\end{center}
{\small {\bf Table 1}\hspace{.5cm}
The charge assignment of extra $U(1)$s which are derived from $E_6$.
These charges are normalized as $\displaystyle\sum_{i\in{\bf 27}}Q_i^2=20$.}
\vspace{.5cm}
\end{figure}
There is also a rank five model called $\eta$-model.
Its charge assignment is also listed in Table 1.
As seen from this table, there is a SM singlet $S$ which has
the coupling $\lambda SH_1H_2$.
$\eta$-model crearly satisfies the above mentioned condition.
On the other hand, in the rank six models this condition imposes rather 
severe constraint on the extra $U(1)_X$ in the low energy region.
In this type of models a right-handed sneutrino $N$ also has 
to get the VEV to break the gauge symmetry into the SM one.
If we try to explain the smallness of the neutrino mass in this
context, $N$ should
get the sufficiently large VEV. In fact, in the case that $N$ has 
a conjugate chiral partner $\bar N$, a sector of $(N, \bar N)$ has a D-flat 
direction and then they can get a large VEV without breaking 
supersymmetry \cite{dflat}. 
This VEV can induce the large right-handed Majorana neutrino mass
through the nonrenormalizable term $(N\bar N)^n/M_{\rm pl}^{2n-3}$
in the superpotential
and then the seesaw mechanism is applicable to yield the small
neutrino mass \cite{sy,sue0}.
However, this usually breaks the direct relation between 
the $\mu$-scale and the mass of the extra neutral gauge boson
because the VEV of $N$ also contributes to the latter.
In order to escape this situation and obtain the extra $U(1)_X$ 
satisfying our condition, 
we need to construct a $U(1)_X$ by taking a linear combination of $U(1)_\psi$
and $U(1)_\chi$ \cite{sy,ms,sue0}.
As such examples, we can construct two low energy extra $U(1)_X$ models.
They are shown in Table 2. The difference between them is the overall
sign.\footnote{As discussed in refs. \cite{sy,sue0}, $Q_{\xi_-}$ can
also be obtained only by changing the field assignments for $Q_\chi$.
This insight allows us to construct new models, which can induce an
interesting neutrino mass matrix \cite{neut} by using the
charge assignments $Q_\chi$ and $Q_{\xi_-}$ for the
different generations \cite{sue0}. However, in this paper we shall not
consider such models for simplicity.}
In these models the right-handed sneutrinos have
no charge of this low energy extra $U(1)_X$.
This is a different situation from the rank five $\eta$-model.
Thus using the D-flat direction of another extra $U(1)$, 
the right-handed sneutrino gets the large VEV which breaks this extra
$U(1)$ symmetry and also can induce the
large Majorana masses for the right-handed neutrinos.
This mechanism may also be related to the inflation of universe and the
baryogenesis as discussed in \cite{infl}.
As a result of this symmetry breaking at the intermediate scale,
only one extra $U(1)_X$ remains as the low energy symmetry.
We will concentrate ourselves on these three $U(1)_X$ models
$(X=\eta,~\xi_\pm)$ in the following study.

We focus our attention to the minimally extended part of
these models with an extra $U(1)_X$ and a SM singlet Higgs $S$. 
Other extra matter fields like color triplet fields 
$(g,\bar g)$ and the right-handed neutrino $N$, which are 
introduced associated with the extension, are irrelevant to the present 
purpose and we can neglect them.
Thus the relevant parts of the superpotential and soft supersymmetry
breaking terms are
\begin{eqnarray}
W&=&\lambda SH_1H_2 + h_UQU^cH_2 + h_DQD^cH_1+h_ELE^cH_1+\cdots, \nonumber\\
{\cal L}_{\rm soft}&=&-\sum_im_i^2|\phi_i|^2 +(A\lambda SH_1H_2+{\rm
h.c}+\cdots)\nonumber\\
&+&{1\over2}(M_W\sum_{a=1}^3\hat\lambda_W^a\hat\lambda_W^a
+M_Y\hat\lambda_Y\hat\lambda_Y+M_X\hat\lambda_X\hat\lambda_X
+M_{YX}\hat\lambda_Y\hat\lambda_X +{\rm h.c.}),
\end{eqnarray}
where $\phi_i$ represents the scalar component of each chiral
superfield contained in the models.
$M_W, M_Y$ and $M_X$ are the gaugino masses.\footnote{
In this expression we introduced the abelian gaugino mass mixing 
as $M_{YX}$, which might exist as the tree level term at the 
Planck scale and also be yielded through the quantum effects.}
We assume the Yukawa coupling $\lambda$ and soft supersymmetry 
breaking parameters to be real, for simplicity.
\begin{figure}[bt]
\begin{center}
\begin{tabular}{|c|c|c||c|}\hline
fields &$SU(3)\times SU(2)$& $Y$ & $Q_{\xi_\pm}$ \\\hline
$Q$  &(3,2)&${1\over 3}$&$\pm {1\over\sqrt6}$ \\    
$U^c$&$(3^\ast,1)$&$-{4\over 3}$&$\pm{1\over\sqrt6}$\\
$D^c$&$(3^\ast,1)$&${2\over 3}$&$\pm{2\over\sqrt 6}$\\
$L$  &(1,2)&$-1$&$\pm {2\over\sqrt 6}$ \\
$E^c$&(1,1)&2&$\pm {1\over\sqrt 6}$\\
$H_1$&(1,2)&$-1$& $\mp {3\over\sqrt 6}$\\
$H_2$&(1,2)&1&$\mp{2\over\sqrt 6}$\\
$g$&(3,1)&$-{2\over 3}$&$\mp {2\over\sqrt 6}$ \\
$\bar g$&$(3^\ast,1)$&${2\over 3}$&$\mp{3\over\sqrt 6}$\\
$S$ &(1,1)&0&$\pm {5\over\sqrt 6}$\\
$N$ &(1,1)&0&0 \\\hline
\end{tabular}\vspace*{.2cm}
\end{center}
{\small {\bf Table 2}\hspace{.5cm}
The charge assignment of the extra $U(1)_X$ which remains unbroken
after the VEV of $N$ becomes nonzero. They are obtained as
$\displaystyle Q_{\xi_\pm}=\pm{\sqrt{15}\over 4}Q_\psi \pm {1\over 4}Q_\chi$.}
\vspace{.5cm}
\end{figure}

\subsection{Abelian gauginos mixing}
Next we briefly review a particular feature in the neutralino sector
caused by the abelian gauge kinetic term mixing in the supersymmetric
multi $U(1)$s models. 
In supersymmetric models gauge fields are extended to vector superfields
\begin{equation}
V_{\rm WZ}(x, \theta, \bar\theta)=-\theta\sigma_\mu\bar\theta V^\mu
+i\theta\theta\bar\theta\bar\lambda-i\bar\theta\bar\theta\theta\lambda
+{1\over 2}\theta\theta\bar\theta\bar\theta D,
\end{equation}
where we used the Wess-Zumino gauge.
A gauge field strength is included in the chiral superfield
constructed from $V_{\rm WZ}$ in the well-known procedure,
\begin{eqnarray}
W_\alpha(x, \theta)&=&(\bar D\bar D)D_\alpha V_{\rm WZ} \nonumber \\
&=&4i\lambda_\alpha-4\theta_\alpha
D+4i\theta^\beta\sigma_{\nu\alpha\dot\beta}
\sigma^{\dot\beta}_{\mu\beta}(\partial^\mu V^\nu-\partial^\nu V^\mu)
-4\theta\theta\sigma_{\mu\alpha\dot\beta}\partial^\mu\bar\lambda^{\dot\beta}.
\end{eqnarray}
Here we should note that $W_\alpha$ of the abelian gauge group is
gauge invariant itself.
In terms of these superfields the supersymmetric gauge invariant 
Lagrangian can be written as
\begin{equation}
{\cal L}={1\over 32}\left(W^\alpha W_\alpha\right)_F
+\left(\Phi^\dagger \exp(2g_0QV_{\rm WZ})\Phi\right)_D, 
\end{equation}
where $\Phi=(\phi, \psi, F)$ is the chiral superfield and represents 
matter fields.
Its generalization to the multi $U(1)$s case is straightforward.
The supersymmetric gauge kinetic parts are obtained by using chiral 
superfields $W_\alpha^a$ and $W^b_\alpha$ for $U(1)_a\times U(1)_b$ as
\begin{equation}
{1\over 32}\left(\hat W^{a\alpha} \hat W^a_\alpha\right)_F
+{1\over 32}\left(\hat W^{b\alpha} \hat W^b_\alpha\right)_F
+{\sin\chi\over 16}\left(\hat W^{a\alpha} \hat W_{\alpha}^b\right)_F.
\end{equation}
Here we introduced the mixing term between the different $U(1)$s.
This can be canonically diagonarized by using the transformation,
\begin{equation}
\left(\begin{array}{c} \hat W^a \\ W^b \\ \end{array}\right)
=\left(\begin{array}{cc}1 & -\tan\chi \\ 0 & 1/\cos\chi \\ 
\end{array}\right)\left(\begin{array}{c} W^a \\ W^b \\
\end{array}\right).
\end{equation}
This transformation affects not only the gauge
field sector but also the sector of gauginos $\lambda_{a,b}$ and 
auxiliary fields $D_{a,b}$.\footnote{
This shift in the D-term changes the scalar potential and can affect the
symmetry breaking at the weak scale. However,
we will not refer to this problem here.} 
As easily seen from the form of the last term in eq. (4),
the change induced in the interactions of gauginos with other fields
through this transformation can be summarized as
\begin{equation}
g_a^0Q_a\hat \lambda^a+g_b^0Q_b\hat
\lambda^b=g_aQ_a\lambda^a+\left(g_{ab}Q_a+g_bQ_b\right)\lambda^b,
\end{equation}
where $\lambda_{a,b}$ are canonically normalized gauginos.
The charges of $U(1)_a$ and $U(1)_b$ are represented by $Q_a$ and
$Q_b$.
The couplings $g_a, g_{ab}$ and $g_b$ are related to the original ones
$g_a^0$ and $g_b^0$ as,
\begin{equation}
g_a=g_a^0, \qquad g_{ab}=-g_a^0\tan\chi, \qquad g_b={g_b^0\over\cos\chi}. 
\end{equation}
These coupling constants at weak scale will be determined by using the 
renormalization group equations from the initial values at the high energy
scale \cite{ms,acq}. However, such a study is beyond our present purpose and 
we will treat them as the parameters in the later analysis.

\subsection{Neutral gauge sector}  
In the previously introduced extra $U(1)_X$ models, 
the gauge symmetry of the electroweak sector at the low energy region 
is $SU(2)_L\times U(1)_Y\times U(1)_X$.
In order to bring the correct symmetry breaking for these models,
we assume that Higgs fields get the VEVs as follows,
\begin{equation}
\langle H_1\rangle=\left(\begin{array}{c}v_1\\0\end{array}\right),
\quad \langle H_2
\rangle=\left(\begin{array}{c}0\\v_2\end{array}\right),
\quad \langle S\rangle=u, 
\end{equation}
where $v_1^2+v_2^2=(246~{\rm GeV})^2(\equiv v^2)$ is assumed.
For simplicity, all VEVs are assumed to be real.
Under these settings in order to constrain the parameters of the
models, we investigate some features of the gauge boson 
sector.

For this purpose we need to determine the physical states at and below the
weak scale \cite{mixing2}. 
The mass mixing between two neutral gauge fields appears associated 
with the spontaneous symmetry breaking due to the VEVs of eq. (9)
around the weak scale.
In the present models the charged gauge sector is the same as that 
of the MSSM. In the neutral gauge sector we introduce 
the Weinberg angle $\theta_W$ in the usual
way,\footnote{
In the following we use the abbreviated notation
$s_W\equiv \sin\theta_W$ and $c_W\equiv \cos\theta_W$.}
\begin{equation}
Z_\mu=\cos\theta_WW_\mu^3-\sin\theta_WB_\mu, \quad
A_\mu=\sin\theta_WW_\mu^3+\cos\theta_WB_\mu.
\end{equation}
Here we used the canonically normalized basis $(Z_\mu, X_\mu)$ so
that $A_\mu$ has already been decoupled from $(Z_\mu, X_\mu)$.
The mass matrix of the neutral gauge fields $(Z_\mu, X_\mu)$ 
can be written as
\begin{equation}
\left( \begin{array}{cc}
m_Y^2 & m^2_{YX} \\
m^2_{YX} & m^2_X \\
\end{array} \right),
\end{equation}
where each element is expressed as
\begin{eqnarray}
&&m_Y^2=m_Z^2, \nonumber \\
&&m_{YX}^2=m_Z^2s_W\tan\chi+{\Delta m^2 \over \cos\chi}, \nonumber \\
&&m_X^2=m_Z^2s^2_W\tan^2\chi 
+2\Delta m^2s_W{\sin\chi \over\cos^2\chi}
+{M_{Z^\prime}^2\over \cos^2\chi}.
\end{eqnarray}
In this expression $m_Z^2, \Delta m^2$ and $M_{Z^\prime}^2$ represent 
the values of corresponding components in case of no 
kinetic term mixing $(\chi=0)$,
\begin{eqnarray}
&&m_Z^2={1\over 2}(g^2_W+g_Y^2)v^2, \nonumber \\
&&\Delta m^2={1 \over 2}(g_W^2+g_Y^2)^{1/2}g_X v^2
\left(Q_1\cos^2\beta -Q_2\sin^2\beta\right), \nonumber \\
&&M_{Z^\prime}^2={1\over 2}g_X^2\left(Q_1^2v_1^2+Q_2^2v_2^2+Q_S^2
u^2\right).
\end{eqnarray}
The mass matrix eq. (11) can be diagonalized by introducing a mixing
angle $\xi$. The canonically normalized mass eigenstates are written
by using $\chi$ and $\xi$. Their concrete expressions are given in
Appendix A. We also present there the interaction Lagrangian of the
neutral gauge bosons and matter fermions for the later use. 

The mixing angle introduced for the diagonalization of
the mass matrix eq. (11) is given by
\begin{equation}
\tan2\xi={-2\cos\chi(m_Z^2s_W\sin\chi+\Delta m^2) \over
M_{Z^\prime}^2+2\Delta m^2s_W\sin\chi
+m_Z^2s^2_W\sin^2\chi-m_Z^2\cos^2\chi}.
\end{equation}
In general the mixing angle $\xi$ is severely constrained to be small
enough by the precise measurements at LEP \cite{exp}.
From the study of radiative symmetry breaking it has been known that
$\tan\beta~{^>_\sim}~1$ is generally favored. In fact,
it has been shown in ref. \cite{sy} that the suitable radiative symmetry
breaking could occur for $1.4 ~{^<_\sim}~\tan\beta~{^<_\sim}~2.1$
in $\xi_-$ model.
We will adopt 
\begin{equation}
\tan\beta \sim 1.5
\end{equation}
as its typical value throughout this paper.
Therefore, in case of $\sin\chi=0$,
since $Q_1\cos^2\beta\simeq Q_2\sin^2\beta$ is not satisfied in the
present three models, we need to consider the possibility that the
small $\xi$
is realized because of $\Delta m^2 \ll M_{Z^\prime}^2$, which
is equivalent to $v_1^2, v_2^2 \ll u^2$.
If $\sin\chi\not=0$, however, there may be a new possibility to
satisfy the smallness of $\xi$ even if $\Delta m^2\ll M_{Z^\prime}^2$
is not satisfied. Such a situation can be expected to occur if the following
condition
\begin{equation}
\sin\chi\sim -{\Delta m^2\over m_Z^2s_W}= 
{g_X\over (g_W^2+g_Y^2)^{1/2}s_W}(Q_1\cos^2\beta-Q_2\sin^2\beta)
\end{equation} 
is valid. 
In this case $Q_1\cos^2\beta\simeq Q_2\sin^2\beta$ is not required 
unlike the $\sin\chi=0$ case but instead of that the tuning of $\sin\chi$ 
becomes necessary. The constraint on the value of $u$ also becomes
very weak.
Since this possibility for the small $m_{YX}^2$ compared with $m_Z^2$ 
is interesting enough for the explanation of the smallness of 
$\vert\xi\vert$,
we will also consider the case with such a mixing angle $\sin\chi$ in the 
following discussion as one of the typical examples.

The present model-independent bound on the mixing angle $\xi$ is
$\vert \xi\vert < 0.01$ \cite{mixxi}. If we impose this bound on the
models, we can restrict
the allowed $u$ range in each model. Here it should be noted that
the mixing angle $\xi$ has no $\lambda$ dependence. 
In order to show this constraint coming from the neutral gauge sector,
we plot the contours of the mixing angle $\vert\xi\vert=0.01$ for each 
model in the $(\sin\chi, \vert u\vert)$ plane in Fig. 1. The lower
regions of the contours are forbidden in each model. 
It is noticeable that rather small value of $\vert u\vert$ is generally 
allowed in $\xi_\pm$-models in comparison with $\eta$-model.
From this figure we find that 
the kinetic term mixing $\sin\chi$ can affect the lower bound 
of $\vert u\vert$ substantially.
In $\eta$-model the larger $\sin\chi$ reduces the required bound of 
$\vert u\vert$ values.
In $\xi_\pm$-models\footnote{$\xi_\pm$-models have the
symmetric feature mutually with respect to the sign of $\sin\chi$ so 
that they are expected to show the similar behavior in their
phenomenology. 
This comes from their characteristics of the charge assignments.}
there are special values of $\sin\chi$ which 
make the lower bound of $\vert u\vert$ very small, as anticipated in
eq. (16).
Thus in these models the rather light extra $Z^0$ may be
possible.\footnote{
In this case the extra $U(1)_X$ gaugino is also expected to affect largely 
the rare phenomena like $\mu\rightarrow e\gamma$ and the EDM of 
an electron \cite{sue}.}

Related to the fact that rather large $\vert u\vert$ is generally
required except for the case with the special $\sin\chi$ value,
it will be useful to remind again the origin of $\mu$-scale in the
present models.
In these models the vacuum expectation value $u$ 
is relevant to the $\mu$-scale.
Based on this feature we may need to put the upper bound on $\lambda$ 
to keep $\mu$ to be the suitable scale from the viewpoint of 
radiative symmetry breaking as discussed in \cite{sy}. 
If we use the present Higgs mass bounds, however, $\lambda$ can be
effectively constrained as shown in the following subsection.

\subsection{Higgs sector}
Higgs sector is changed from that of the MSSM due to the 
existence of the singlet $S$ and its coupling $\lambda SH_1H_2$
to the Higgs doublets $H_1$ and $H_2$.
Its brief study can give us some useful informations of the allowed
region of the parameter space\cite{hs,drees}.
If we take account of the abelian gauge kinetic term mixing,
the scalar potential for the Higgs sector can be written as,
\begin{eqnarray}
V&=&{1 \over 8}\left(g_W^2+g_Y^2\right)\left(v_1^2+v_2^2\right)^2\nonumber\\
&&+{1 \over 8}\left\{g_Y\tan\chi(v_1^2-v_2^2)+{g_X \over \cos\chi}
\left( Q_1v_1^2+Q_2v_2^2+Q_Su^2)\right)\right\}^2 \nonumber \\
&&+\lambda^2v_1^2v_2^2+\lambda^2u^2v_1^2+\lambda^2u^2v_2^2 
+m_1^2v_1^2 +m_2^2v_2^2+m_S^2u^2 - 2A\lambda uv_1v_2,
\end{eqnarray}
where $Q_1, Q_2$ and $Q_S$ represent the extra $U(1)_X$ charges of Higgs
chiral superfields $H_1, H_2$ and $S$.
At the minimum of this potential,
the mass matrices for the Higgs sector are given as follows respectively,\\
$\bullet$ charged Higgs scalar sector:\hfill
\begin{equation}
\left[ {1 \over 2}m_Z^2c_W^2\sin 2\beta
\left(1-{2\lambda^2 \over g_W^2}\right)+A\lambda u\right]
\left( \begin{array}{cc}
\tan\beta & 1 \\
1&\cot\beta \\
\end{array}\right),
\end{equation}
$\bullet$ neutral Higgs scalar sector:\hfill
\scriptsize
$$
\left( \begin{array}{ccc}
m_Z^2\cos^2\beta(1+{\zeta_1^2\over \tilde g^2})+A\lambda u\tan\beta&
{m_Z^2\over 2}\sin 2\beta(-1+{\zeta_1\zeta_2+4\lambda^2\over
\tilde g^2})-A\lambda u & 
{m_Z\cos\beta\over \tilde g}\left(u
{\zeta_1\zeta_3+4\lambda^2\over \sqrt 2}- \sqrt 2A\lambda\tan\beta\right)\\
{m_Z^2\over 2}\sin 2\beta(-1+{\zeta_1\zeta_2+4\lambda^2\over
\tilde g^2})-A\lambda u&
m_Z^2\sin^2\beta(1+{\zeta_2^2\over \tilde g^2})+A\lambda u\cot\beta&
{m_Z\sin\beta\over \tilde g}\left(u
{\zeta_2\zeta_3+4\lambda^2\over \sqrt 2}- \sqrt 2A\lambda\cot\beta\right) \\
{m_Z\cos\beta\over \tilde g}\left(u{\zeta_1\zeta_3+
4\lambda^2\over \sqrt 2}- \sqrt 2A\lambda\tan\beta\right)&
{m_Z\sin\beta\over \tilde g}\left(u
{\zeta_2\zeta_3+4\lambda^2\over \sqrt 2}- \sqrt 2A\lambda\cot\beta\right)&
{1\over 2}\zeta_3^2u^2+{A\lambda \over u}{m_Z^2 \over
\tilde g^2}\sin 2\beta\\
\end{array}\right) ,
$$
\normalsize
\begin{flushright}
(19)
\end{flushright}
\normalsize
where $\tilde g=\sqrt{g_W^2+g_Y^2}$ and we define $\zeta_1, \zeta_2$ 
and $\zeta_3$ as\footnote{It may be 
useful to note that the sign of $\zeta_1, \zeta_2$ and $\zeta_3$ is reversed
between $\xi_+$-model with $\sin\chi$ and $\xi_-$-model with $-\sin\chi$.}
\setcounter{equation}{19}
\begin{equation}
\zeta_1=g_Y\tan\chi+{g_XQ_1\over\cos\chi}, \quad 
\zeta_2=-g_Y\tan\chi+{g_XQ_2\over\cos\chi},\quad
\zeta_3={g_XQ_S\over\cos\chi}.
\end{equation}

The overall factor of a mass matrix of the charged Higgs sector 
is somehow  changed 
from that of the MSSM due to the coupling $\lambda SH_1H_2$.
However, the mass eigenstate of charged Higgs scalar can be obtained 
in the same form as the MSSM case
\begin{equation}
H^{\pm}=\sin\beta H_1^\pm +\cos\beta H_2^{\mp\ast},
\end{equation}
and its mass eigenvalues are expressed as
\begin{equation}
M_{H^\pm}^2=m_Z^2c_W^2\left(1 -{2\lambda^2 \over g_W^2}\right)
+{A\lambda u \over \sin\beta\cos\beta}.
\end{equation}
The $\lambda^2$ term is added to the MSSM one and then the charged Higgs 
mass takes smaller value than that of the MSSM for the same value of
$\mu=\lambda u$.
On the other hand, the neutral Higgs mass matrix is too complex to be
diagonalized analytically.
However, if we note that the smallest eigenvalue of the matrix is smaller 
than the smallest diagonal component,
we can find the tree level upper bound of the lightest neutral Higgs mass.
By diagonalizing the $2\times 2$ submatrix at the upper left corner 
of eq. (19) we can obtain\cite{drees,sy}
\begin{equation}
m_{h^0}^2 \le m_Z^2 \left[\cos^22\beta + {2\lambda^2 \over g_W^2+g_Y^2}
\sin^22\beta + {1 \over g_W^2+g_Y^2}\left(
\zeta_1\cos^2\beta+\zeta_2\sin^2\beta)^2\right)\right].
\label{ubhig}
\end{equation}
The first two terms correspond to the bound which is derived from 
the usually studied model extended with a gauge singlet.

As easily seen from these results, these Higgs masses have the
crucial dependence on $\lambda$ and $u$. 
One of the important difference between the present models and the
MSSM comes from the fact that the $\mu$-term is replaced by the 
Yukawa coupling $\lambda SH_1H_2$.
If we impose the present experimental
bounds on the Higgs masses, useful constraints can be obtained in the
$(\lambda, u)$ plane.
The present mass bounds on both of the charged Higgs and the lightest neutral
Higgs are $\sim$44 GeV \cite{pdata}. 
We use this bound and show the allowed
region in the $(\lambda, u)$ plane in Fig. 2.
Since it is found to be insensitive to the 
models and also the $\sin\chi$ value,
we take $\xi_-$-model with $\sin\chi=0$ as an example.
Here for the lightest neutral Higgs we used the result obtained by the
numerical diagonalization of the mass matrix eq. (19).  
It should be noted that only the $u>0$ region is allowed. This is
completely dependent on our choice $(A>0)$ for the sign 
of $A$.\footnote{ The $A$ and $u$
dependence of the Higgs mass eigenvalues is included in the terms, which are
composed of $Au$ and 
even powers of each of them. Thus the sign of $u$ is related to that of $A$.
Here it should also be noted that in the present notation $u>0$ corresponds 
to the ordinary $\mu<0$ case.}

Additional important constraints on $\mu$ can be obtained from the
condition in the $(\mu, M_W)$ plane coming from the 
search of the neutralinos and charginos at 
LEP \cite{aleph}.
If we assume $\tan\beta \sim 1.5$, the allowed region in this
plane is roughly estimated as,\footnote{It should be noted that 
this restriction has been derived under some assumptions, for example, 
the gaugino unification relation $M_Y={5\over 3}\tan^2\theta_WM_W$.
However, we will apply them for the general $M_Y$ and $M_W$ here.
These constraints correspond to 
the condition for the chargino mass $m_1^c~{^>_\sim}~65$GeV.}
\begin{equation}
\begin{array}{ll}
|\mu|,~M_W ~{^>_\sim}~40~{\rm GeV}\quad & ({\rm for}~ \lambda u >0),\\
|\mu|,~M_W ~{^>_\sim}~100~{\rm GeV}\quad &({\rm for}~ \lambda u <0).\\
\end{array}
\end{equation}
The chargino sector in the present model is not altered from the MSSM and then
these conditions on $\mu$ can be used as the constraint for $\lambda$
and $u$. Thus the allowed region of the $(\lambda, u)$ plane are found to 
be determined by the lower bound of the lightest neutral Higgs mass 
for all models. It corresponds to the surrounded region by the 
dashed lines in Fig. 2.
If we combine this with the result obtained from Fig. 1, we can restrict the
allowed region in the $(\lambda, u)$ plane for each model with a
certain $\sin\chi$ value.
We will use this fact later.

\section{The decay width of $\tilde\chi_2^0$ into $\tilde\chi_1^0$}
\subsection{Neutralino sector}
In this subsection we examine the structure of the neutralino sector and
also define the mass eigenstates of the charginos and squarks/sleptons
sector, which are necessary for the calculation of the neutralino decay.
Starting from the superpotential and soft supersymmetry breaking terms 
given in eq. (1) and using the canonically normalized basis defined 
by eq. (6), 
we can write down the modified
quantities from the MSSM, which are relevant to the neutralino sector, 
that is, the neutralino mass matrix and the gaugino-fermion-sfermion 
interaction terms.
If we take the canonically normalized gaugino basis 
${\cal N}^T=(-i\lambda_W^3, -i\lambda_Y, 
-i\lambda_X, \tilde H_1, \tilde H_2, \tilde S)$ and define
the neutralino mass term as
${\cal L}_{\rm mass}^n=-{1\over 2}{\cal N}^T{\cal MN}+{\rm h.c.}$,
the 6 $\times$ 6 neutralino mass matrix ${\cal M}$ can be expressed as
\begin{equation}
\left( \begin{array}{cccccc}
M_W & 0 & 0 &m_Zc_W\cos\beta & -m_Zc_W\sin\beta &0 \\
0 & M_Y & C_1 & -m_Zs_W\cos\beta & m_Zs_W\sin\beta &0 \\
0 & C_1 & C_2 & C_3 & C_4 & C_5 \\
m_Zc_W\cos\beta &-m_Zs_W\cos\beta &  C_3& 0 & 
\lambda u & \lambda v\sin\beta \\
-m_Zc_W\sin\beta & m_Zs_W\sin\beta &C_4 & \lambda u &
0 & \lambda v\cos\beta \\
0 & 0& C_5 & \lambda v\sin\beta & \lambda v\cos\beta & 0\\
\end{array} \right),
\end{equation}
where $v$ and $u$ are defined by eq. (9).
Matrix elements $C_1\sim C_5$ are components which are affected by the 
kinetic term mixing. They are represented as
\begin{eqnarray}
&&C_1=-M_Y\tan\chi +{M_{YX}\over\cos\chi},\quad
C_2=M_Y\tan^2\chi+{M_X \over
\cos^2\chi}-{2M_{YX}\sin\chi\over\cos^2\chi},  \nonumber \\
&&C_3={1\over \sqrt 2}\left(g_Y\tan\chi+{g_XQ_1\over\cos\chi}\right)
v\cos\beta,\quad
C_4={1\over \sqrt 2}\left(-g_Y\tan\chi+{g_XQ_2\over\cos\chi}\right)
v\sin\beta,\nonumber\\
&&C_5={1\over \sqrt 2}{g_XQ_S\over\cos\chi}u.
\end{eqnarray}
Neutralino mass eigenstates $\tilde\chi_i^0(i=1\sim 6)$ are related 
to ${\cal N}_j$
through the mixing matrix $U_{ij}$ as
\begin{equation}
\tilde\chi^0=U^T{\cal N}.
\end{equation}
The change in the gaugino interactions can be confined into 
the extra $U(1)_X$ gaugino sector and new interaction terms can be 
expressed as,
\begin{eqnarray}
&&{i \over \sqrt 2}\left[\tilde \psi^*\left(-g_YY\tan\chi+{g_XQ_X\over
\cos\chi}\right)\lambda_X\psi-\left(-g_YY\tan\chi+{g_XQ_X\over\cos\chi}
\right)\bar\lambda_X\bar \psi\tilde \psi\right. \nonumber \\
&&\left.+H^*\left(-g_YY\tan\chi+{g_XQ_X\over
\cos\chi}\right)\lambda_X\tilde
H-\left(-g_YY\tan\chi+{g_XQ_X\over\cos\chi}
\right)\bar\lambda_X\bar{\tilde H}H\right]
\end{eqnarray} 
where $\psi$ and $\tilde \psi$ represent quarks/leptons and squarks/sleptons.
Higgs fields $ (H_1, H_2, S)$ are summarized as $H$ and the corresponding
Higgsinos $(\tilde H_1,\tilde H_2,\tilde S)$ are denoted as 
$\tilde H$.
The charges of $U(1)_Y$ and $U(1)_X$ are denoted as $Y$ and $Q_X$. 
As a result, the parts corresponding to the gaugino component of the 
neutralino $\tilde\chi_i^0$-fermion-sfermion vertices 
are represented by the following factors
\begin{eqnarray}
&&Z_i^L(Y, Q_X)=-{1\over \sqrt 2}\left[g_WU_{1i}\tau_3+g_YYU_{2i}+
\left(-g_YY\tan\chi+{g_XQ_X\over \cos\chi}\right)U_{3i}\right], \nonumber \\
&&\overline{Z_i^R}(Y, Q_X)=-{1\over \sqrt 2}\left[g_YYU_{2i}+
\left(-g_YY\tan\chi+{g_XQ_X\over \cos\chi}\right)U_{3i}\right],
\end{eqnarray}
where the suffixes $L$ and $R$ stand for the chirality of the coupled
matter fields $\psi$ and their charges are defined in terms of the 
left-handed chiral basis as presented in Tables 1 and 2.

Additionally, it is also useful to define the chargino and squark mass 
eigenstates here for the forthcoming calculation.
Taking account of eq. (1), the chargino mass terms are given as
\begin{equation}
{\cal L}_{\rm mass}^c=-\left(H_1^-, -i\lambda^-\right)
\left(\begin{array}{cc}-\lambda u& \sqrt 2m_Zc_W\cos\beta\\
\sqrt 2m_Zc_W\sin\beta& M_W \\ \end{array}\right)
\left(\begin{array}{c}H_2^+ \\ -i\lambda^+\\ \end{array}\right).
\end{equation}
The mass eigenstates $\tilde\chi_i^\pm$ are defined in terms of the 
weak interaction eigenstates through the unitary transformations,
\begin{equation}
\left(\begin{array}{c}\tilde \chi_1^+\\ \tilde \chi_2^+\\ \end{array}\right)
\equiv W^{(+)\dagger}\left(\begin{array}{c}H_2^+\\ -i\lambda^+\\
\end{array}\right),  \qquad
\left(\begin{array}{c}\tilde \chi_1^-\\ \tilde \chi_2^-\\ \end{array}\right)
\equiv W^{(-)\dagger}
\left(\begin{array}{c}H_1^-\\ -i\lambda^-\\ \end{array}\right).
\end{equation}

Squarks and sleptons are also relevant to the neutralino decay.
When we consider this subject, all flavors can be treated in the same
way except for the top sector.
If they appear in the internal lines, the stop may be especially 
important because of the largeness of its Yukawa couplings 
and then we only consider the stop sector in such cases.
However, in the neutralino decay modes which contain the
ordinary fermions in the final states, top quark is too heavy to be
included in them and it is irrelevant to such processes.

In the following analysis we do not consider the flavor 
mixing in the squark and slepton sector, for simplicity. 
Thus the sfermion mass
matrices can be reduced into the $2\times 2$ form for each flavor.
This $2\times 2$ sfermion mass matrix can be written in terms of 
the basis $(\tilde f_L, \tilde f_R)$ as
\small
\begin{equation}
\left(\begin{array}{cc} |m_f|^2
+M_L^2+ D_L^2& 
m_f(A_f+\lambda uR_f)\\
m_f^\ast(A_f^\ast+\lambda uR_f)& 
|m_f|^2+M_R^2+ D_R^2\\ 
\end{array}\right),
\end{equation}
\normalsize
where $m_f$ and $M_{L,R}^2$ are the masses of ordinary fermion $f$ and
its superpartners $\tilde f_{L,R}$, respectively. 
We assume $M_{L,R}^2$ is universal for all flavors.
$R_f$ is $\cot\beta $ for up-sector and $\tan\beta$ for down-sector.
Soft supersymmetry breaking
parameters $A_f$ are the dimensionful coefficients of three scalar
partners of the corresponding Yukawa couplings.
$D_L^2$ and $D_R^2$ represent the $D$-term contributions, which are 
modified in the present models as follows
\begin{eqnarray}
&&D_L^2=\pm{1\over 2}m_Z^2\cos 2\beta(1-(1\pm Y)s_W^2)+{1\over
4}g_X^2Q_X^\prime (Q_1^\prime v_1^2 +Q_2^\prime v_2^2+Q_S^\prime u^2), 
\nonumber \\
&&D_R^2=-{1\over 2}m_Z^2s_W^2Y\cos 2\beta +{1\over
4}g_X^2Q_X^\prime (Q_1^\prime v_1^2 +Q_2^\prime v_2^2+Q_S^\prime u^2), 
\end{eqnarray}
where the upper sign in $D_L$ corresponds to the up-sector sfermions and the
lower one to down-sector sfermions.
The primed charge $Q^\prime_X$ stands for the modified charge due to the 
kinetic term mixing and defined as 
$g_XQ_X^\prime=-g_YY\tan\chi+g_XQ_X/\cos\chi$.
We should note that these $D$-term contributions cannot be neglected in the 
extra $U(1)_X$ where $u$ tends to be large. In such cases 
it will be useful to note that the positivity condition of 
the sfermion masses may induce no condition on the soft scalar masses.
We define the mass eigenstates $(\tilde f_1, \tilde f_2)$ as
\begin{equation}
\left(\begin{array}{c} \tilde f_1 \\ \tilde f_2\\ \end{array}\right)
\equiv V^{f\dagger}\left(\begin{array}{c}\tilde f_L \\ \tilde f_R\\ 
\end{array}\right).
\end{equation}
Under our assumption for the reality of soft SUSY parameters 
the above chargino and sfermion mass matrices are  
real and then $W^{(\pm)}$ and $V^f$ become the orthogonal matrices.

By now we have finished the preparations for the calculation of neutralino
decay in the present models. 
If R-parity is conserved and the lightest neutralino is the lightest
superparticle, the decay of the next-to-lightest neutralino
$\tilde\chi_2^0$ into the lightest neutralino $\tilde\chi_1^0$ can
be expected to appear in the various superparticle decay processes. 
As the representative decay modes of $\tilde\chi_2^0$ into
$\tilde\chi_1^0$, the tree level three body decay
$\tilde\chi_2^0\rightarrow\tilde\chi_1^0f\bar f$ and the one-loop
radiative decay $\tilde\chi_2^0\rightarrow\tilde\chi_1^0\gamma$ have
been calculated in the MSSM framework \cite{mssmdecay,am,ndecay}.
In these studies, which decay mode of these becomes dominant has been 
shown to be crucially dependent on the composition of $\tilde\chi_2^0$ and
$\tilde\chi_1^0$ and then on the SUSY parameters.
It is very interesting that one-loop decay mode can easily dominate
the tree level process in the suitable parameter region.
As was recently stressed in ref. \cite{cdf}, if the CDF type 
event relevant to $\tilde\chi_2^0\rightarrow\tilde\chi_1^0\gamma$
happen to be observed dominantly instead of $\tilde\chi_2^0\rightarrow 
\tilde\chi_1^0f\bar f$, it can constrain the SUSY parameter space severely.
In the following part of this section we shall analyze the decay widths of
$\tilde\chi_2^0\rightarrow\tilde\chi_1^0f\bar f$ and
$\tilde\chi_2^0\rightarrow\tilde\chi_1^0\gamma$ in the present extra
$U(1)_X$ models and also qualitatively discuss the condition 
on the SUSY parameters
for the $\tilde\chi_2^0\rightarrow\tilde\chi_1^0\gamma$
dominance.

There exist other decay modes like
two body decay into the lightest Higgs $\tilde\chi_2^0\rightarrow 
\tilde\chi_1^0 h^0$ and the cascade decay mediated through the
chargino as
$\tilde\chi_2^0\rightarrow \tilde\chi_1^+(e\bar\nu_e)\rightarrow
\tilde\chi_1^0\bar e\nu_e(e\bar\nu_e)$. If the $h^0$ is light enough
for the threshold to be opened satisfying $m_{\tilde\chi_2^0}-
m_{\tilde\chi_1^0} >m_{h^0}$, the first one can be a relevant mode.
The second one may not be suppressed if $\tilde\chi_1^+$ is 
lighter than $\tilde\chi_2^0$ even in the case that $\tilde\chi_2^0$ 
is composed of the same ingredients as the case where 
$\tilde\chi_2^0\rightarrow\tilde\chi_1^0f\bar f$ is suppressed.  
Although these points should be taken into account in the analysis,
through the numerical calculation of the mass eigenvalues 
$h^0$ at least seems to be heavy enough not to open the threshold 
in the parameter
region $(\lambda u,m_Z)$ where we are interested in.
For the chargino mediated cascade decay the threshold can be opened
but the existence of its suppression mechanism has been pointed out in
ref. \cite{am}. Therefore,
in this paper we concentrate our attention on the comparison of
$\tilde\chi_2^0\rightarrow\tilde\chi_1^0\gamma$ and 
$\tilde\chi_2^0\rightarrow\tilde\chi_1^0f\bar f$.
For this purpose we shall firstly calculate the decay width of both modes.
We are particularly interested in the case of rather small neutralino masses
since in such a case these neutralino decays may be observed in the experiment
in near future.

\subsection{$\tilde\chi_2^0\rightarrow\tilde\chi_1^0f\bar f$}
There are two types of diagrams which contribute to the tree
level three body decay. They are shown in Fig.3.
Top quark cannot be a final state so that the contribution from
diagram (b) is generally suppressed by the small Yukawa coupling. 
The phase space integral can be analytically done
in the limit that the mass of the final state fermion $f$ is zero.
This seems to be generally rather good approximation and we adopt
this result of the phase space integral in the present estimation.
Thus the decay width for this process can be expressed as\footnote{
It should be noted that in the limit of $m_f\rightarrow 0$
there is no interference term like 
$F_{f_L}^{(1)}F_{f_R}^{(1)}$ between the different fermion 
chiralities in ${\cal F}_\alpha^2~ (\alpha=1,2)$.}
\begin{equation}
\Gamma(\tilde\chi_j^0\rightarrow\tilde\chi_i^0f\bar f)=
{1\over 96\pi^3}\left[{m_j^4-m_i^4 \over m_j^3}
\left(m_j^4+m_i^4-8m_j^2m_i^2\right)+24m_jm_i^4\ln{m_j\over m_i}\right]
\sum_{\alpha=1}^4 {\cal F}^2_\alpha,
\end{equation}
where the vertex factors ${\cal F}_\alpha$ can be expressed by using the 
mixing matrix element $U_{ij}$ in the neutralino sector as
\begin{eqnarray}
&&{\cal F}_1= {1\over 4m_{Z_1}^2}\left[\left(g^{(1)}
+{g_X\over 2\cos\chi}Q_1\sin\xi\right)
U_{4j}U_{4i}+\left(-g^{(1)}+{g_X\over
2\cos\chi}Q_2\sin\xi\right)U_{5j}U_{5i}\right. \nonumber\\
&&\hspace*{7cm}\left.+ {g_X\over 2\cos\chi}Q_S\sin\xi U_{6j}U_{6i}\right]
\left(F_{f_L}^{(1)}+ F_{f_R}^{(1)}\right), \nonumber \\
&&{\cal F}_2= {1\over 4m_{Z_2}^2}\left[\left(g^{(2)}+{g_X\over 2\cos\chi}Q_1
\cos\xi\right)U_{4j}U_{4i}+\left(-g^{(2)}
+{g_X\over 2\cos\chi}Q_2\cos\xi\right)U_{5j}U_{5i}\right.\nonumber\\
&&\hspace*{7cm}\left.+ {g_X\over 2\cos\chi}Q_S\cos\xi U_{6j}U_{6j}\right]
\left(F_{f_L}^{(2)}+F_{f_R}^{(2)}\right),\nonumber \\
&&{\cal F}_3^{(f=U)}=\sum_{\alpha=1,2}
{1\over 8M_{\tilde f_\alpha}^2}\left[\left(
Z^L_{1j}(Y,Q_X)U_{5i}+
Z^L_{1i}(Y,Q_X)U_{5j}\right)h_fV_{1\alpha}^{f2}\right.
\nonumber \\
&&\hspace*{2.5cm}-\left( \overline{Z^R_{1j}}(Y,Q_X)U_{5i}
+\overline{Z^R_{1i}}(Y,Q_X)U_{5j}\right)h_fV_{2\alpha}^{f2}
-2h_f^2U_{5j}U_{5i}V^f_{2\alpha}V^f_{1\alpha} \nonumber\\
&&\hspace*{2.5cm}\left.+\left(Z^L_{1j}(Y,Q_X)
\overline{Z^R_{1i}}(Y,Q_X)+Z^L_{1i}(Y,Q_X)
\overline{Z^R_{1j}}(Y,Q_X)\right)V^{f}_{1\alpha}V^f_{2\alpha}
\right],
\end{eqnarray}
where 
\begin{eqnarray}
&&g^{(1)}={g_W\over 2c_W}\left(\cos\xi+s_W\tan\chi\sin\xi\right), 
\nonumber\\
&&g^{(2)}={g_W\over 2c_W}\left(-\sin\xi+s_W\tan\chi\cos\xi\right).
\end{eqnarray}
${\cal F}_1$ and ${\cal F}_2$ comes from the diagram (a).
The mass eigenvalues of the neutral gauge bosons are expressed as
$m_{Z_1}$ and $m_{Z_2}$.
$M_{\tilde f_\alpha}^2$ is mass eigenvalues of sfermion mass matrix 
eq. (32). 
The effective neutral current couplings $F_{f_L}^{(1)}$ {\it etc.} 
are deviated from ones of the MSSM 
due to the existence of the extra $U(1)_X$ and the abelian gauge kinetic 
term mixing. Their concrete expressions are presented in Appendix A.
$Z^L_i(Y,Q_X)$ and $\overline{Z^R_i}(Y,Q_X)$ are defined by eq. (29).
The diagram (b) gives ${\cal F}_{3,4}^f$ and
${\cal F}_4^{(f=D,E)}$ is obtained by replacing $Z^L_{1j}(Y,Q_X)$,
$\overline{Z^R_{1j}}(Y,Q_X)$ and $U_{5j}$ in ${\cal F}_3^{(f=U)}$
with $Z^L_{2j}(Y,Q_X)$,
$\overline{Z^R_{2j}}(Y,Q_X)$ and $U_{4j}$,
respectively. 

It is useful to examine under what condition this decay width can be
suppressed based on eqs. (35) and (36). 
As was noticed up to now \cite{am}, there can happen the dynamical suppression
depending on the composition of $\tilde\chi_2^0$ and $\tilde\chi_1^0$
which is determined by the SUSY parameters. For the contribution from 
${\cal F}_1$ and ${\cal F}_2$
they are suppressed unless both of $\tilde\chi_2^0$ and
$\tilde\chi_1^0$ are dominated by the Higgsinos.
In ${\cal F}_3$ and ${\cal F}_4$ there are the contributions 
from both of the 
Higgsino and gaugino components in $\tilde\chi_2^0$ and
$\tilde\chi_1^0$ and then it seems to be difficult to expect the 
suppression due to the neutralino composition.
However, there is the crucial suppression due to the small Yukawa
coupling and also the small left-right mixing $V_{12}^f$ in the
sfermion mass matrix.
These features can be summarized as follows. There appears the dynamical
suppression effectively in such a case that one of $\tilde\chi_2^0$ and  
$\tilde\chi_1^0$ is dominated by gauginos and the other is dominated
by Higgsinos.
Although this is the same as the MSSM situation, there is a noticeable
feature in the present extra $U(1)_X$ models.
In case of the $\tilde S$ dominated neutralino, it has no mixings
with $\lambda_W$ and $\lambda_Y$. Moreover, it has no couplings with 
ordinary fermions. 
If this is the case,  it is not necessary for the
gaugino dominated neutralino to be an almost pure photino in order to 
suppress this three body decay unlike the MSSM.
Later this point will be discussed in more detail again.
  
\subsection{$\tilde\chi_2^0\rightarrow\tilde\chi_1^0\gamma$}
Next we proceed to the calculation of the one-loop radiative decay
$\tilde\chi_2^0\rightarrow\tilde\chi_1^0\gamma$. This has already been
studied in the MSSM framework \cite{mssmdecay}.  
From the gauge invariance, as suggested in \cite{effect}, it is easily found 
that the effective interaction describing this process is given as
\begin{equation}
{\cal L}_{\rm eff}={\cal G}\bar{\tilde\chi_j^0}
\sigma_{\mu\nu}\tilde\chi_i^0F^{\mu\nu}. 
\end{equation}
Using this effective coupling ${\cal G}$, the decay width is written as
\begin{equation}
\Gamma(\tilde\chi_j^0\rightarrow\tilde\chi_i^0\gamma)=
{\vert {\cal G}\vert^2(m_j^2-m_i^2)^3\over 2\pi m_j^3},
\end{equation}
where $m_i$ and $m_j$ are the masses of $\tilde\chi_i^0$ and
$\tilde\chi_j^0$.
Our main problem is the estimation of the effective coupling ${\cal G}$.
One-loop diagrams contributing this coupling are given in Fig. 4.
In the diagrams (1a) and (1b), only the stop contribution cannot be neglected 
because of its large Yukawa coupling.
After some algebraic manipulation,
it is obvious that this coupling can be obtained as the coefficient
of $q\hspace{-2mm}/\cdot\epsilon\hspace{-1.5mm}/$ terms where $q_\mu$ and 
$\epsilon_\mu$ are the momentum and the polarization vector of photon.
In case of small neutralino masses $m_{i,j} \ll m, M$ where $m$ and $M$ 
respectively represent masses of fermions and bosons in the internal
lines, the neutralino mass dependence disappears from these one-loop 
amplitudes.  
Its only dependence on the neutralino sector comes through the mixing
matrix $U_{ij}$ of the neutralino sector. 
The effective coupling ${\cal G}$ can be summarized as follows,
\begin{eqnarray}
&&{\cal G}=-{e\over 32\pi^2}\left[\sum_{\alpha=1,2}
{3\over m_t}f({M_{\tilde t_\alpha}^2\over m_t^2})G_1^\alpha
+\sum_{\alpha=1,2}{1\over m_\alpha}
\left\{f({M_{H^\pm}^2\over m_\alpha^2})G_2^\alpha
+f({M_W^2\over m_\alpha^2})G_3^\alpha\right.\right. \nonumber \\
&&\hspace*{2cm}\left.\left.-\left(4I({M_W^2\over m_\alpha^2})+3J({M_W^2\over
m_\alpha^2})\right)G_4^\alpha
+{M_W\over 2m_\alpha}J({M_W^2\over m_\alpha^2})G_5^\alpha\right\}\right], 
\end{eqnarray}
where $m_{\alpha}$ and $M_{H^\pm}$ stand for the masses of the charginos
 and the charged Higgs. 
The charged Higgs mass expression is presented in eq. (22).
The first and second summations should be taken for the stop mass 
eigenstates and the chargino mass eigenstates, respectively. 
Each term with a vertex factor $G_i^\alpha$ comes from
Feynman diagram numbered with $i$ in Fig. 4 and their concrete expressions 
are presented in Appendix B.
Kinematical functions $f(r), I(r)$ and $J(r)$ are defined as
\begin{eqnarray}
&&f(r)={1 \over 1-r}\left[1+{r \over 1-r}\ln r\right],\\
&&I(r)={1 \over 2(1-r)^2}\left[1+r+{2r \over 1-r}\ln r\right],\\
&&J(r)={1 \over 2(1-r)^2}\left[-3+r-{2 \over 1-r}\ln r\right].
\end{eqnarray}
For checking this formula, we assume that $\lambda u, M_W, M_Y, 
M_X \ll m_Z$ and
the stop mass matrix is diagonal ($V_{\alpha\beta}=\delta_{\alpha\beta}$).  
In such a case, for $W^{(\pm)}_{\alpha\beta}$, 
the situation is the same as the MSSM
and they can be taken as,\footnote{
Here the sign conventions are taken so as to
make both mass eigenvalues positive.}
\begin{eqnarray}
&&W_{12}^{(+)}=-W_{21}^{(+)}=W_{11}^{(-)}=W_{22}^{(-)}=1, \nonumber \\
&&W_{11}^{(+)}=W_{22}^{(+)}=W_{12}^{(-)}=W_{21}^{(-)}=0.
\end{eqnarray}
Mass eigenvalues of charginos are approximately written as
\begin{equation}
m^{c}_1=\sqrt 2m_Zc_W\cos\beta, \qquad
m^{c}_2=\sqrt 2m_Zc_W\sin\beta.
\end{equation}
For $U_{ij}$, if we put $g_X=0$ and $\lambda\rightarrow0$ 
but keeping $\mu(=\lambda u)$  constant, $U_{ij}$ can be approximated as
\begin{eqnarray}
&&U_{1i}=s_W, \quad U_{2i}=c_W, \quad U_{4j}=\sin\beta, \quad
 U_{5j}=\cos\beta,\quad {\rm other}~ U_{ij}=0, \nonumber\\
&&Z_i^L(Y)=-\sqrt 2g_2s_WQ_{\rm em}, \quad 
\overline{Z^R_i}(Y)=\sqrt 2g_2s_WQ_{\rm em},
\quad Z_j^L(Y)=\overline{Z_j^R}(Y)=0.
\end{eqnarray}
Using these expressions, it can be easily checked that ${\cal G}$ is reduced 
to the MSSM result
calculated in this parameter setting \cite{mssmdecay}.

The feature of eq. (40) is rather similar to the one of the MSSM.
As easily seen from the structure of $G_i^\alpha$ in Appendix B, 
there is not the special neutralino configuration in which the drastic 
suppression mechanism works for 
$\Gamma(\tilde\chi_j^0\rightarrow\tilde\chi_i^0\gamma)$
unlike $\Gamma(\tilde\chi_j^0\rightarrow\tilde\chi_i^0\bar ff)$. 
This is an important feature to consider the neutralino decay processes. 

\subsection{The radiative decay dominant condition}
As was clarified through the study of the CDF event 
$ee\gamma\gamma+/\hspace{-2.7mm}E_T$ \cite{cdf},
the neutralino decay can give the valuable information on the
SUSY parameters.
Based on the naive perturbative sense, 
as $\tilde\chi_2^0\rightarrow\tilde\chi_1^0\gamma$ is the higher 
order process compared with
$\tilde\chi_2^0\rightarrow\tilde\chi_1^0f\bar f$, the former is
expected to be largely suppressed by the small couplings 
compared with the latter.\footnote{It has been suggested 
that there is also a kinematical suppression of the three body decay 
when $\tilde\chi_2^0$ and $\tilde\chi_1^0$ are nearly degenerate 
$m_j^2-m_i^2 \ll m_j^2$ \cite{am}. However, in our study we will not 
refer to such a parameter region.} 
However, in the present case the neutralinos are complicatedly composed of 
the various ingredients and two decay modes imply the different feature
depending on their compositions which are determined by the SUSY parameters.
If the signature of the radiative decay mode is dominantly observed,
the SUSY parameter space can be strictly restricted due to the suppression
condition of the tree level three body decay.
Thus it will be useful to study how this situation can be changed
in the extra $U(1)_X$ models. 

For this investigation it is convenient to rewrite the neutralino 
mass matrix eq. (25) in terms of the usual photino and Higgsino 
basis which is often used in the MSSM case.
It can be written as

\scriptsize
$$ \left( \begin{array}{cccccc}
M_Ws_W^2+M_Yc_W^2 & (M_W-M_Y)s_Wc_W & c_WC_1 & 0 & 0 &0 \\
(M_W-M_Y)s_Wc_W & M_Wc_W^2+M_Ys_W^2 & -s_WC_1 & m_Z &0 &0 \\
c_WC_1 & -s_WC_1 & C_2 & C_3\cos\beta-C_4\sin\beta & 
C_3\sin\beta+C_4\cos\beta & C_5 \\
0 & m_Z &C_3\cos\beta-C_4\sin\beta & -\lambda u\sin 2\beta & 
\lambda u\cos 2\beta & 0 \\
0 & 0 &C_3\sin\beta+C_4\cos\beta & \lambda u\cos 2\beta &
\lambda u\sin 2\beta & \lambda v \\
0 & 0& C_5 & 0 & \lambda v & 0\\
\end{array} \right), $$

\normalsize
\begin{flushright}
(47)
\end{flushright}
where we define the neutralino basis of this matrix as
$(-i\lambda_1,-i\lambda_2,-i\lambda_3,,\tilde H_a,\tilde H_b,\tilde H_c)$. 
Throughout this study we assume that the gaugino masses $M_W$ and $M_Y$ take
the smaller value than 200GeV. 

In the MSSM case the radiative decay dominant condition is expressed 
as \cite{cdf,am}
\setcounter{equation}{47}
\begin{equation}
M_W \simeq M_Y, \qquad \tan\beta \simeq 1.
\end{equation} 
The second one is natural from the viewpoint of radiative
symmetry breaking and we assume that it is satisfied in our study
as mentioned before. The first one is nontrivial but 
it may not be necessarily required strictly in some parameter region
as pointed out in ref. \cite{am}.   
As easily seen from the part of eq. (47) corresponding to the MSSM 
neutralino sector, 
we find that in the MSSM with the condition eq. (48) the almost pure 
photino $\lambda_1$
and the one of Higgsinos $\tilde H_b$ become 
the lower two neutralino mass eigenstates as far as
$M_W, M_Y, \lambda u \ll m_Z$. 
This situation realizes the suppression of the three body decay
as discussed in the last part of subsection 3.2.
On the other hand, this kind of suppression of three body decay 
seems not to be realized in the present extra $U(1)_X$ models even 
if the above condition is satisfied.
This is because of the existence of the extra $U(1)_X$ gaugino which has
the mixings with every neutralino components.
Thus in order to suppress the tree level three body decay 
it is necessary to resolve this mixing effectively and produce the
purely Higgsino-type neutralino.
Although various possibilities may be considered, we are particularly
interested in the case with $M_W\not\simeq M_Y$.

The first possibility is to make $\lambda_1$ and/or $\lambda_2$ decouple from 
one of the Higgsinos by imposing 
\begin{equation}
C_1\simeq 0,\qquad u \gg v,
\end{equation}
in addition to eq. (48). The first one requires $M_Y\sin\chi=M_{YX}$ 
and it is always satisfied in case of no kinetic term mixing.
The second one should be usually satisfied in the extra $U(1)_X$
models to overcome the small mixing condition on $\xi$ as
discussed in the previous part.
As shown in Tables 1 and 2, $C_3\sin\beta \pm C_4\cos\beta\simeq 0$ 
cannot be satisfied in the present extra
$U(1)_X$ models. 
However, if $u$ is large enough, $C_5$ becomes large
and as a result $C_3\sin\beta \pm C_4\cos\beta\simeq 0$ 
can be effectively satisfied.
Under this situation the Higginos $\tilde H_b$ can decouple from 
$\lambda_1$ and $\lambda_2$.
The value of $\lambda$ is related to which neutralinos become the
lower two neutralino mass eigenstates and then it seems not to be severely
restricted by requiring the radiative decay dominace.
As easily seen from the above mass matrix, $\lambda_3$ ans $\tilde
H_c$ tend to decouple from other fields under the condition (49) 
and the situation is reduced to the MSSM one. 
The feature of $\tilde\chi_2^0$ and $\tilde\chi_1^0$ is expected
to be similar to the one of the MSSM. 
When the composition of these states is interchanged, the same
suppression is also expected to occur.
In this possiblity it should be noted that 
$M_W \sim M_Y$ will not be necessarily required like the MSSM
as far as $M_X$ takes the similar value as $M_W$ and $M_Y$.
In $\xi_{\pm}$-model with the suitable $\sin\chi$ value,
the large $u$ is not necessarily needed. In such a case, although the
$Z^\prime$ becomes rather light, the radiative decay dominance
can not be expected.
In this case $\sin\chi=0$ seems to be preferable for the
radiative decay dominance.

The second possibility is to make the lightest 
neutralino be the almost pure $\tilde S$. 
As mentioned in the sebsection 3.2, $\tilde S$ has no mixings with
$\lambda_W$ and $\lambda_Y$ and also no couplings with ordinary fermions.
Thus if we consider the situation that the next-to-lightest neutralino 
is the mixture of $\lambda_W$ and $\lambda_Y$ and the lightest
neutralino is dominated by $\tilde S$, the three body decay can be suppressed.
This gives a new window which does not require the condition 
$M_W\sim M_Y$.
A very light neutralino dominated by $\tilde S$ is considered in the
different context in ref. \cite{ce}. 
To realize this situation it is necessary to impose
\begin{equation}
C_2 \gg C_5, \qquad C_1 \simeq 0, \qquad \lambda u > m_Z.
\end{equation}
The first one means that $M_X$ needs to be rather large compared with
$u$. We need a particular supersymmetry breaking mechanism which can
realize the large hierarchy among soft gaugino masses such as $M_X \gg M_Y$.
If $u \gg v$ which is generally the preferable situation for 
the extra $U(1)_X$ models, the next-to-lightest neutralino is almost 
the mixture of $\lambda_1$ and $\lambda_2$({\it i.e.} $\lambda_W$ and 
$\lambda_Y$) and
also the lightest neutralino $\tilde H_c$ which is purely $\tilde S$.
Starting from this case, we can get other composition for the lightest
neutralino which realize the radiative decay dominace by shifting 
the values of $M_X$ and $u$. 
If we assume $u~{^>_\sim}~ v$, the lightest neutralino
becomes the mixture of $\tilde H_b$ and $\tilde H_c$.
This situation can be realized in the $\xi_{\pm}$-models with the suitable 
$\sin\chi$ value as found from Fig.1.
If the condition $C_2 \gg C_5$ is changed into $C_5 \gg C_2 \gg v$
which is equivalent to $u \gg M_X \gg v$,
the lightest neutralino becomes $\tilde H_b$ and the situation
becomes similar to the MSSM case except that $\tilde\chi_2^0$ 
does not need to be a photino like state but is enough to be any states 
composed of $\lambda_W$ and $\lambda_Y$. 
It should be noted that these new possibilities are related to
the large $\mu (>m_Z)$ and/or $\sin\chi\not=0$ case, where
$\lambda_X$ and $\tilde S$ can play the crucial role.
In the $\sin\chi\not=0$ case, $C_1 \simeq 0$ requires the existence of 
nonzero $M_{YX}$. The validity of this condition should be checked by
using the RGE study in each model.

\section{Numerical analysis}
The arguments in the previous section are qualitative one on 
the suppression mechanism for the three body decay of $\tilde\chi_2^0$
 compared with the radiative decay. 
It is necessary to proceed the numerical calculations to treat the
subtlety of the parameter dependences and also restrict in more
quantitative way the SUSY 
parameter space where the radiative neutralino decay becomes the 
dominant mode.
As suggested above, there may be a new window of the SUSY parameters 
in the present
extra $U(1)_X$ models and it may be possible to escape the constraint
eq. (48) on the gaugino mass in the MSSM. 
To clarify this we compare the two decay modes numerically. 
In the study of this direction the most interesting parameters are 
the gaugino masses.
In addition to them, $u$ and $\lambda$ will be also important
in the present models because it is relevant to the extra $Z^0$ mass
and also the $\mu$-scale. 

Before going to the numerical analysis of these decay widths, 
it will be useful to summarize the allowed parameter region.
We have already presented the contraints on $\lambda$ and $u$
in Figs. 1 and 2.
By combining these results, for the typical values of 
$\sin\chi$ the allowed region of $u$ is roughly estimated as,
\begin{eqnarray}
\sin\chi=0 &&\left\{
\begin{array}{ll} 
\eta{\rm -model}:&\quad u~{^>_\sim}~1375{\rm GeV},\qquad
0.1~{^<_\sim}~\lambda~{^<_\sim}~0.42\\
\xi_+{\rm -model}:&\quad u~{^>_\sim}~ 550{\rm GeV},\qquad
0.1~{^<_\sim}~\lambda~{^<_\sim}~0.53\\
\xi_-{\rm -model}:&\quad u~{^>_\sim}~550{\rm GeV},\qquad
0.1~{^<_\sim}~\lambda~{^<_\sim}~0.53\\
\end{array}\right. \nonumber \\
\sin\chi=0.2&&   \left\{
\begin{array}{ll} 
\eta{\rm -model}:&\quad u~{^>_\sim}~1200{\rm GeV},\qquad
0.1~{^<_\sim}~\lambda~{^<_\sim}~0.42\\
\xi_+{\rm -model}:&\quad u~{^>_\sim}~ 775{\rm GeV},\qquad
0.1~{^<_\sim}~\lambda~{^<_\sim}~0.52\\
\xi_-{\rm -model}:&\quad u~{^>_\sim}~200{\rm GeV},\qquad
0.1~{^<_\sim}~\lambda~{^<_\sim}~0.66\\
\end{array}\right. \nonumber \\
\sin\chi=-0.2 &&  \left\{ 
\begin{array}{lll}  
\eta{\rm -model}:&\quad u~{^>_\sim}~1525{\rm GeV},\qquad
0.1~{^<_\sim}~\lambda~{^<_\sim}~0.42\\
\xi_+{\rm -model}:&\quad u~{^>_\sim}~ 200{\rm GeV},\qquad
0.1~{^<_\sim}~\lambda~{^<_\sim}~0.66\\
\xi_-{\rm -model}:&\quad u~{^>_\sim}~775{\rm GeV},\qquad
0.1~{^<_\sim}~\lambda~{^<_\sim}~0.52\\
\end{array}\right. \nonumber \\
\end{eqnarray} 
where $M_W ~{^>_\sim}~40~{\rm GeV}$
should be satisfied.
Here we should note that $\sin\chi$ affects the neutralino decay
widths eqs. (35) and (39) not only directly through the vertex factors
and the mixing matrix but also indirectly through determining the
lower bound of $u$.
For the soft supersymmetry breaking parameters we assume
the typical values as follows,
\begin{equation}
A=A_f=200~{\rm GeV}, \qquad 
M_L=M_R=200~{\rm GeV}.
\end{equation} 
Additionally,
$g_Y=g_X$ and $M_{YX}=0$ are also assumed.\footnote{Although these should be
determined in terms of RGEs analysis, we make these assumptions 
only for simplicity.}
The gaugino mass $M_X$ is treated as a free parameter and also
the gaugino masses $M_W$ and $M_Y$ are assumed to take not so large
values such as $40{\rm GeV}~{^<_\sim}~M_W,M_Y~{^<_\sim}~200$GeV.
Under this parameter setting, the branching ratio
$Br\equiv \Gamma(\tilde\chi_2^0\rightarrow\tilde\chi_1^0\gamma)/
[\Gamma(\tilde\chi_2^0\rightarrow\tilde\chi_1^0f\bar f)+
\Gamma(\tilde\chi_2^0\rightarrow\tilde\chi_1^0\gamma)]$ 
are studied in the $(\lambda, u)$ and $(M_W, M_Y)$ planes 
for the typical values of $\sin\chi$ and $M_X$.
Through this study we found that the decay width of the radiative decay
is in the rather wide range $O(10^{-6}\sim 10^{-10})$GeV depending on
the parameters.
Although from the viewpoint of the experimental detectability 
it may be possible to
restrict furthermore the parameter region based on the absolute value of
$\Gamma(\tilde\chi_2^0\rightarrow\tilde\chi_1^0\gamma)$,
we are interested mainly in the radiative decay dominance conditions 
and then we focus our 
attention only on the $Br$ value here.
It should also be noted that $Br$ gives the same value for $\xi_-$-model with 
$\sin\chi$ and $\xi_+$-model with $-\sin\chi$.

At first we examine $Br$ under the condition of
$M_W=M_Y=M_X(~{^<_\sim}~200$GeV) in the $(\lambda, u)$ plane.
As an example, we take $\xi_-$-model which has the rather small lower
bound of $u$. In this model it is expected that there is no severe
restriction on the value of $\lambda$.
In fact the numerical studies show that $Br>0.98$ is realized 
almost through all the region which satisfies the constraints
coming from Figs.1 and 2, 
although for a certain $\mu$ value around $\sim 600$
GeV there is a shallow valley
where $Br$ gives the slightly smaller value compared with other region.
That valley moves in the $(\lambda, u)$ plane by the order of
$\lambda u\sim O(10^{1\sim 2})$GeV following the change of the 
value of $\sin\chi$ from $-0.2$ to 0.2.
This shift originated from the change of $\sin\chi$ becomes larger as $M_X$  
becoming larger. When $M_X$ becomes larger,
$Br<0.90$ occurs at the small $\lambda$ region such as $\lambda~{^<_\sim}~
0.2$.
These qualitative features are found to be common to all models.
The difference between $\eta$-model and $\xi_\pm$-model is that the
latter can have the smaller bound of $u$. As a result, for the same
value of $\lambda$, $\mu$ in $\xi_\pm$-models can take the smaller
values than that in $\eta$-model. In such a small $\mu$ region $Br$ has the 
tendency to become smaller as far as the small gaugino masses are assumed.
This is because the gaugino-Higgsino mixing cannot be extracted in the 
lower lying neutraino eigenstates.
Anyway we can safely conclude that the radiative decay dominance is
good enough in the whole region of $(\lambda, u)$ as far as 
$M_W=M_Y=M_X$ is satisfied.

Next we proceed to the study of $M_W$ and $M_Y$ dependence of $Br$.
For this purpose we estimate $Br$ in the $(M_W, M_Y)$ plane.
In Fig.5 we show the results for $\xi_-$-model as an example.
The global feature of this kind of plot seems to be characterized by
the value of $\mu(=\lambda u)$ if $M_X$ is fixed.
In the case of $\lambda u~{^<_\sim}~m_Z$ (Figs. 5a, 5b and 5c), 
$M_W\simeq M_Y$ seems not to be severely required. 
This point has been already pointed out in the MSSM case \cite{am}.
However, in this model the larger violation of the relation 
$M_W\simeq M_Y$ seems to be allowed compared with the MSSM case.
When $M_X$ becomes larger compared with $M_W$ and $M_Y$, the $Br>0.9$
region shrinks into the smaller $M_W, M_Y$ region and also there
appears the new $Br>0.9$ region in the large $M_W, M_Y$ domain,
where $M_W\sim M_Y$ is not required.
These behavior of $Br$ may be understood as follows.
Accompanied with the change of $M_X$, the level crossing occurs
between $\tilde\chi_1^0$ and $\tilde\chi_2^0$ and then their
ingredients are interchanged. And in the region of $M_X$ where the
separation between  $\tilde\chi_1^0$ and $\tilde\chi_2^0$ is large enough, 
$Br >0.9$ is realized.
In the case of $\lambda u > m_Z$ (Figs. 5d, 5e and 5f), 
when $M_X$ is smaller compared 
with $m_Z$, the $Br>0.9$ region appears as the beltlike zone
around the $M_W\sim M_Y$ line but the width of this region is not so narrow.
This means that the next-to-lightest neutralino should be the almost
photino $\lambda_1$ to realize the radiative decay dominance
and then $M_W\simeq M_Y$ is preferable.
Under this condition the mixture of $\lambda_2,\lambda_3,\tilde H_a,
\tilde H_b$ and $\tilde H_c$ can decouple from $\lambda_1$.
As $M_X$ becomes larger, the $Br>0.9$ region has the tendency to occupy
the wider space where $M_W\simeq M_Y$ is not required.
The reason of this $Br$ behavior can be understood from the qualitative 
arguments in the previous section.
Although we show here the results only for one model,
we have checked that other models also showed the similar qualitative 
features. So these results can be considered as qualitatively 
general ones.

Finally we would like to stress that in the extra $U(1)_X$ models 
there is the wide parameter region 
where the radiative decay becomes the dominant mode of the neutralino decay.
This region contains the new possibility such that the relation 
$M_W\sim M_Y$ is completely violated in comparison with the
corresponding parameter space to the case of the MSSM \cite{am}.
This can be possible because of the existence of $\lambda_X$ and
$\tilde S$.
The neutralino decay may give us various informations on the extra
gauge structure.

\section{Summary}
We studied the decay of the next-to-lightest neutralino into
the lightest neutralino in the extended models with an extra $U(1)_X$
and a SM singlet Higgs $S$, which can solve the $\mu$-problem as the 
result of its radiative symmetry breaking. 
In this study we took account of the abelian gaugino kinetic term
mixing. 
At first we investigated the neutral gauge sector and Higgs sector
in order to constrain the parameter space of the models.
Through this analysis we showed that the VEV $\langle S\rangle$ and
the Yukawa coupling $\lambda$ of the singlet Higgs $S$ were constrained 
in the suitable region.
Next the width of the one-loop radiative decay and the tree level
three body decay were calculated. 
Based on those results the suppression condition of the three body decay
were qualitatively discussed and we suggested that there could be a new 
possibility to escape the constraint on the gaugino masses $M_W\simeq M_Y$ 
for the realization of such a suppression in the MSSM.
This is due to the existence of the extra $U(1)_X$ gaugino and the 
singlet field $S$. 
For more quantitative analysis the branching ratio of the radiative
decay was numerically estimated in the $(\lambda, u)$ and $(M_W,
M_Y)$ planes.
As a result we found that the $\mu$-problem solvable extension with 
the extra $U(1)_X$ could largely modify the parameter space which
realize the radiative decay dominance from that of 
the MSSM. Especially, it was pointed out that the condition $M_W \simeq M_Y$ 
for the gaugino masses is not necessarily required for the 
radiative decay dominance as far as $M_X$ is large enough.
In the extra $U(1)_X$ models the slepton and squark decays which contain
the above processes as the subprocesses can be largely affected by the 
existence of the extra gauge bosons and the singlet Higgs.
These results seem to be interesting for the future accelerator experiments.
In the supersymmetric models the extension with extra $U(1)$s
may have the interesting and fruitful phenomena in their superpartner 
sector and its extra gauge structure may be seen through the study of the 
superpartner sector.
The further study of this aspect will be worthy enough.
\vspace{1cm}

\noindent
{\Large\bf Acknowledgement}\vspace{.3cm}\\
The author would like to thank J.~Kubo for useful discussions. 
This work is partially supported by a Grant-in -aid for Scientific
Research from the Ministry of Education, Science and Culture(\#08640362).

\newpage
\noindent
{\Large\bf Appendix A}
\vspace{.3cm}\\
In this appendix we give the concrete expressions of the interaction 
Lagrangian of the neutral gauge sector.
Original states which are not canonically normalized
are represented by the mass eigenstates $({\cal A}^\mu, 
Z_1^\mu, Z_2^\mu)$ as
\begin{eqnarray}
&&\hat A^\mu={\cal A}^\mu -c_W\tan\chi\left(\sin\xi Z_1^\mu+\cos\xi
Z_2^\mu\right), \nonumber \\
&&\hat Z^\mu=\left(\cos\xi+s_W\tan\chi\sin\xi\right)Z_1^\mu
+\left(-\sin\xi+s_W\tan\chi\cos\xi\right)Z_2^\mu, \nonumber \\
&&\hat X^\mu={\sin\xi\over\cos\chi} Z_1^\mu 
+{\cos\xi\over \cos\chi} Z_2^\mu,
\end{eqnarray}
where ${\cal A}^\mu$ stands for the real photon field and $Z_1^\mu$ is
understood as $Z^{\mu}$ observed at LEP.
Using these mass eigenstates, the interaction terms of these gauge fields
with ordinary quarks and leptons in this model can be expressed as,
\begin{eqnarray}
&&{\cal L}_{\rm int}=J_\mu^{\rm em}{\cal A}^\mu
+j_{\mu}^{(1)}Z_1^\mu+j_{\mu}^{(2)}Z_2^\mu, \nonumber \\
&&\hspace*{6mm}j_{\mu}^{(1)}=F_{f_L}^{(1)}\bar f_L\gamma_\mu f_L+
F_{f_R}^{(1)}\bar f_R\gamma_\mu f_R, \nonumber \\
&&\hspace*{6mm}j_{\mu}^{(2)}=F_{f_L}^{(2)}\bar f_L\gamma_\mu f_L+
F_{f_R}^{(2)}\bar f_R\gamma_\mu f_R, 
\end{eqnarray}
where the coefficients $F_{f_L}^{(1)}$ {\it etc.} are defined as
\begin{eqnarray}
&&F_{f_L}^{(1)}=\left(\tau^3-2Q_{\rm em}s^2_W\right)g^{(1)}
-eQ_{\rm em}c_W\tan\chi\sin\xi
+{g_X\over 2\cos\chi}Q_X^{f_L}\sin\xi, \nonumber\\
&&F_{f_R}^{(1)}= -2Q_{\rm em}s^2_Wg^{(1)}
-eQ_{\rm em}c_W\tan\chi\sin\xi
+{g_X\over 2\cos\chi}Q_X^{f_R}\sin\xi, \nonumber\\
&&F_{f_L}^{(2)}=\left(\tau^3 -2Q_{\rm em}s^2_W\right)g^{(2)}
-eQ_{\rm em}c_W\tan\chi\cos\xi
+{g_X\over 2\cos\chi}Q_X^{f_L}\cos\xi, \nonumber\\
&&F_{f_R}^{(2)}= -2Q_{\rm em}s^2_Wg^{(2)}
-eQ_{\rm em}c_W\tan\chi\cos\xi
+{g_X\over 2\cos\chi}Q_X^{f_R}\cos\xi. 
\end{eqnarray}
$Q_X^{f_L}$ and $Q_X^{f_R}$ stand for the $U(1)_X$ charges of
$f_L$ and $f_R$.

\newpage
\noindent
{\Large\bf Appendix B}
\vspace{.3cm}\\
We give here the concrete expressions of the vertex factors 
$G_i^\alpha (i=1\sim 5)$ in eq. (40).
\small
\begin{eqnarray}
&&\hspace*{-.8cm}G_1^\alpha=-{2\over 3}\left[V_{1\alpha}^\dagger 
V_{2\alpha}^\dagger
\left(\bar{Z_j^{R}}(-{4\over 3})Z_{1i}^L({1\over 3})
-\bar{Z_i^{R}}(-{4\over3})Z_{1j}^L({1\over 3})\right)
\right.\nonumber \\
&&\hspace*{.3cm}\left.-h_UV_{1\alpha}^\dagger V_{1\alpha} 
\left(U_{5j}Z_{1i}^L({1\over 3})
-U_{5i}Z_{1j}^L({1\over3})\right)-h_UV_{2\alpha}^\dagger V_{2\alpha}
\left(U_{5i}\overline{Z_j^R}(-{4\over 3})
-U_{5j}\overline{Z_i^R}(-{4\over3})\right)\right],\\
&&\hspace*{-.8cm}G_2^\alpha={\sin 2\beta\over 2}\left[
g_W^2W_{2\alpha}^{(+)}W_{2\alpha}^{(-)}\left(U_{4i}U_{5j}
-U_{5i}U_{4j}\right)+W_{1\alpha}^{(+)}W_{1\alpha}^{(-)}
\left(Z_{2i}^L(-1)Z_{1j}^L(1)-Z_{2j}^L(-1)Z_{1i}^L(1)\right)\right.
\nonumber\\
&&\hspace*{0cm}\left.-g_WW_{2\alpha}^{(+)}W_{1\alpha}^{(-)}
\left(U_{5j}Z_{2i}^L(-1)-U_{5i}Z_{2j}^L(-1)\right)
-g_WW_{2\alpha}^{(-)}W_{1\alpha}^{(+)}
\left(U_{4i}Z_{1j}^L(1)-U_{4j}Z_{1i}^L(1)\right)\right]\nonumber\\
&&\hspace*{0cm}+\lambda\sin^2\beta\left[g_WW_{2\alpha}^{(-)}
W_{1\alpha}^{(+)}\left(U_{4j}U_{6i}-U_{6j}U_{4i}\right)+
W_{1\alpha}^{(+)}W_{1\alpha}^{(-)}\left(Z_{2i}^L(-1)U_{6j}
-Z_{2j}^L(-1)U_{6i}\right)\right]\nonumber \\
&&\hspace*{0cm}+\lambda\cos^2\beta\left[g_WW_{2\alpha}^{(+)}
W_{1\alpha}^{(-)}\left(U_{5i}U_{6j}-U_{6i}U_{5j}\right)+
W_{1\alpha}^{(+)}W_{1\alpha}^{(-)}\left(Z_{1j}^L(1)U_{6i}
-Z_{1i}^L(1)U_{6j}\right)\right],\\
&&\hspace*{-.8cm}G_4^\alpha=-{g_W^2 \over\sqrt 2}
\left[-W_{2\alpha}^{(+)\dagger}
W_{1\alpha}^{(-)\dagger}\left(U_{1i}U_{4j}-U_{4i}U_{1j}\right)
+W_{1\alpha}^{(+)\dagger}W_{2\alpha}^{(-)\dagger}
\left(U_{5i}U_{1j}-U_{1i}U_{5j}\right)\right.\nonumber\\
&&\hspace*{7cm}\left.+{1\over \sqrt 2} W_{1\alpha}^{(+)\dagger}
W_{1\alpha}^{(-)\dagger}\left(U_{5i}U_{4j}-U_{4i}U_{5j}\right)\right],\\
&&\hspace*{-.8cm}G_5^\alpha=g_W\left[\cos\beta\left(
g_WW_{2\alpha}^{(-)\dagger}W_{2\alpha}^{(-)}
\left(U_{1i}U_{4j}-U_{4i}U_{1j}\right)
\right.\right.\nonumber \\
&&\hspace*{.5cm}+{1\over \sqrt 2}W_{1\alpha}^{(-)\dagger}W_{1\alpha}^{(-)}
\left(U_{4j}Z_{2i}^L(-1)-U_{4i}Z_{2j}^L(-1)\right)
-{\lambda\over \sqrt 2}W_{1\alpha}^{(+)\dagger}W_{1\alpha}^{(+)}
\left(U_{5j}U_{6i}-U_{6j}U_{5i}\right)\nonumber\\
&&\hspace*{.5cm}\left.+W_{1\alpha}^{(-)}W_{2\alpha}^{(-)\dagger}
\left(U_{1j}Z_{2i}^L(-1)-U_{1i}Z_{2j}^L(-1)\right)
-\lambda W_{1\alpha}^{(+)}W_{2\alpha}^{(+)\dagger}
\left(U_{1j}U_{6i}-U_{6j}U_{1i}\right)\right)\nonumber\\
&&\hspace*{0.1cm}+\sin\beta\left(-g_WW_{2\alpha}^{(+)\dagger}W_{2\alpha}^{(+)}
\left(U_{1i}U_{5j}
-U_{5i}U_{1j}\right)\right. \nonumber \\
&&\hspace*{.5cm}+{1\over \sqrt 2}W_{1\alpha}^{(+)\dagger}W_{1\alpha}^{(+)}
\left(U_{5j}Z_{1i}^L(1)-U_{5i}Z_{1j}^L(1)\right)
-{\lambda\over \sqrt 2}
W_{1\alpha}^{(-)\dagger}W_{1\alpha}^{(-)}
\left(U_{4j}U_{6i}-U_{6j}U_{4i}\right)\nonumber\\
&&\hspace*{0.5cm}\left.\left.-W_{1\alpha}^{(+)}W_{2\alpha}^{(+)\dagger}
\left(U_{1j}Z_{1i}^L(1)-U_{1i}Z_{1j}^L(1)\right)
-\lambda W_{1\alpha}^{(-)}W_{2\alpha}^{(-)\dagger}
\left(U_{1j}U_{6i}-U_{6j}U_{1i}\right)\right)\right],
\end{eqnarray}
\normalsize
where in these equations we abbreviate the $U(1)_X$ charges in
the expression of $Z^L_i(Y,Q_X)$ and $\overline{Z^R_i}(Y,Q_X)$.
$G_3^\alpha$ can be obtained by making the replacement such as
$\sin\beta\rightarrow \cos\beta$ and $\cos\beta\rightarrow -\sin\beta$
in $G_2^\alpha$.

\newpage
\noindent
{\Large\bf Figure Captions}
\vspace{.4cm}\\
{\Large\bf Fig. 1} 
\vspace{.2cm}\\
The allowed region in the $(\sin\chi,\vert u\vert)$ plane due to
the constraint on the mixing angle $\xi$ between the extra $U(1)_X$ 
and the ordinary $Z^0$. 
The contours of $\vert\xi\vert=0.01$ for three models are
drawn. $\xi_-$, $\xi_+$ and $\eta$-models correspond to
solid, dashed and dotted-dashed lines, respectively.
The lower region of each contour is forbidden. 
\vspace{.4cm}\\
{\Large\bf Fig. 2} 
\vspace{.2cm}\\
The allowed region in the $(\lambda, u)$ plane for $\xi_-$-model with 
$\sin\chi=0$.
The contours of the present mass bounds of the lightest neutral Higgs
scalar and the charged Higgs scalar are shown by the dashed and
dashed-dotted lines, respectively.
The surrounded region by the dashed lines and
the upper region of the dashed-dotted one are allowed.
The solid lines represent the boundary $(A:~\lambda u=40{\rm GeV}$ and 
$B:~\lambda u=-100{\rm GeV})$ coming from the experimental searches of 
charginos and neutralinos. The region sandwiched between them is forbidden. 
\vspace{.4cm}\\
{\Large\bf Fig. 3} 
\vspace{.2cm}\\
Diagrams contributing to the tree level three body decay
$\tilde\chi_j^0\rightarrow\tilde\chi_i^0f\bar f$. 
\vspace{.4cm}\\
{\Large\bf Fig. 4} 
\vspace{.2cm}\\
One-loop diagrams contributing to 
$\tilde\chi_j^0\rightarrow\tilde\chi_i^0\gamma$.
The chirality flip occurs at the fermion internal lines and/or Yukawa
vertices. In Figs. (2a) and (2b) we show the representative ones. 
\vspace{.4cm}\\
{\Large\bf Fig. 5a} 
\vspace{.2cm}\\
The contours of the branching ratio $Br=0.9, 0.7$ and $0.5$ of $\xi_-$-model 
with $\sin\chi=0$ in the $(M_W, M_Y)$ plane, which are represented 
by solid, dashed and dashed-dotted lines, respectively.
Parameters are set as $\lambda=0.15$, $u=600$GeV and
$M_X=50$GeV.
\vspace{.4cm}\\  
{\Large\bf Fig. 5b}
\vspace{.2cm}\\ 
The same contours of $Br$ as Fig.5a.
Parameters are set as $\lambda=0.15$, $u=600$GeV and $M_X=400$GeV.
\vspace{.4cm}\\ 
{\Large\bf Fig. 5c}
\vspace{.2cm}\\ 
The same contours of $Br$ as Fig.5a.
Parameters are set as $\lambda=0.15$, $u=600$GeV and $M_X=1000$GeV.
\vspace{.4cm}\\ 
{\Large\bf Fig. 5d}
\vspace{.2cm}\\ 
The samw contours of $Br$ as Fig.a.
Parameters are set as $\lambda=0.5$, $u=600$GeV and $M_X=50$GeV.
\vspace{.4cm}\\ 
{\Large\bf Fig. 5e} 
\vspace{.2cm}\\ 
The same contours of $Br$ as Fig.6a.
Parameters are set as $\lambda=0.5$, $u=600$GeV and $M_X=400$GeV.
\vspace{.4cm}\\ 
{\Large\bf Fig. 5f} 
\vspace{.2cm}\\ 
The same contours of $Br$ as Fig.6a.
Parameters are set as $\lambda=0.5$, $u=600$GeV and $M_X=1000$GeV.

\newpage
\pagestyle{empty}
\begin{figure}
\begin{center}
\epsfxsize=\textwidth
\epsfbox{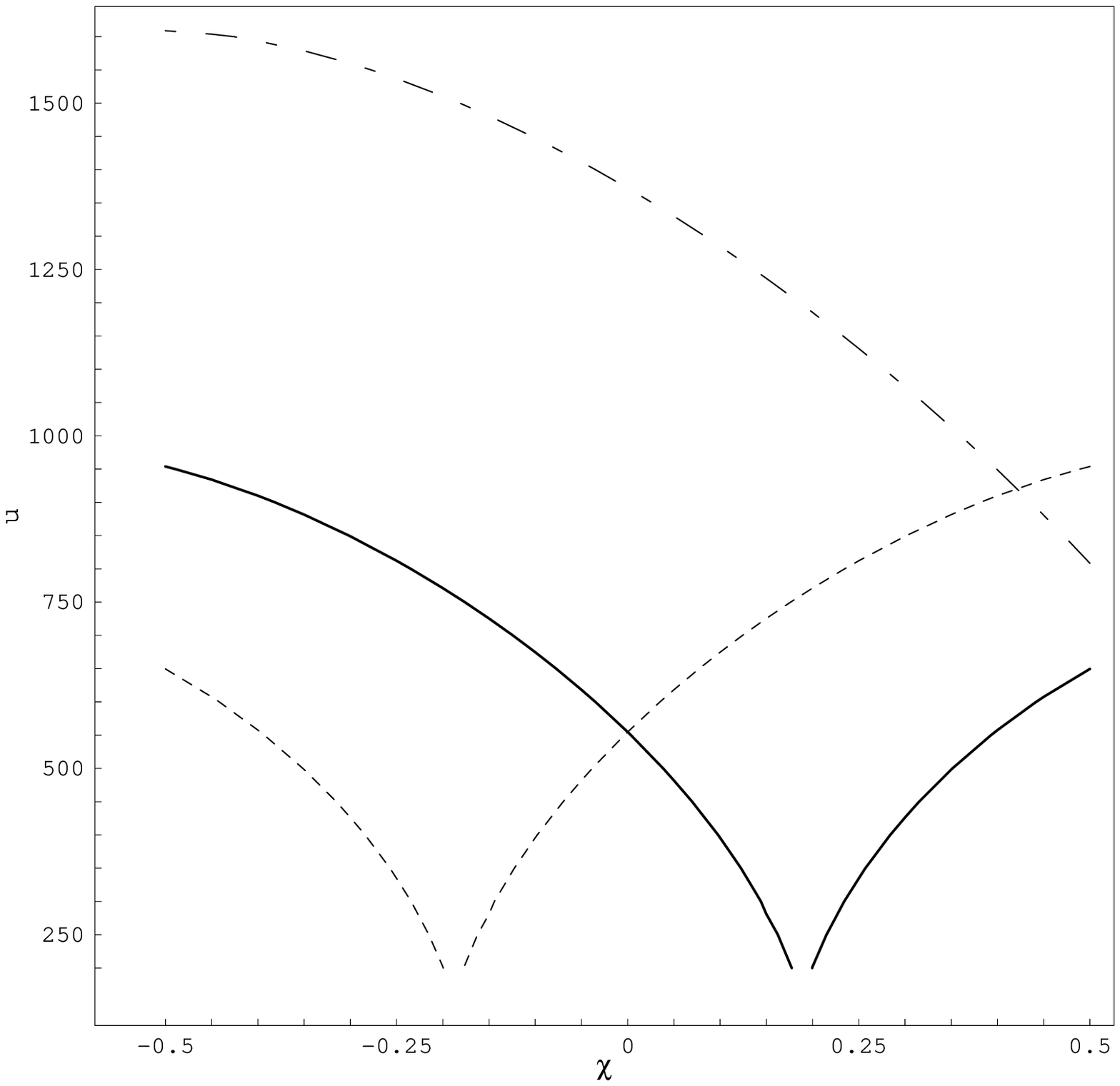}

{\Large\bf Fig. 1}
\end{center}
\end{figure}

\newpage
\begin{figure}
\begin{center}
\epsfxsize=\textwidth
\epsfbox{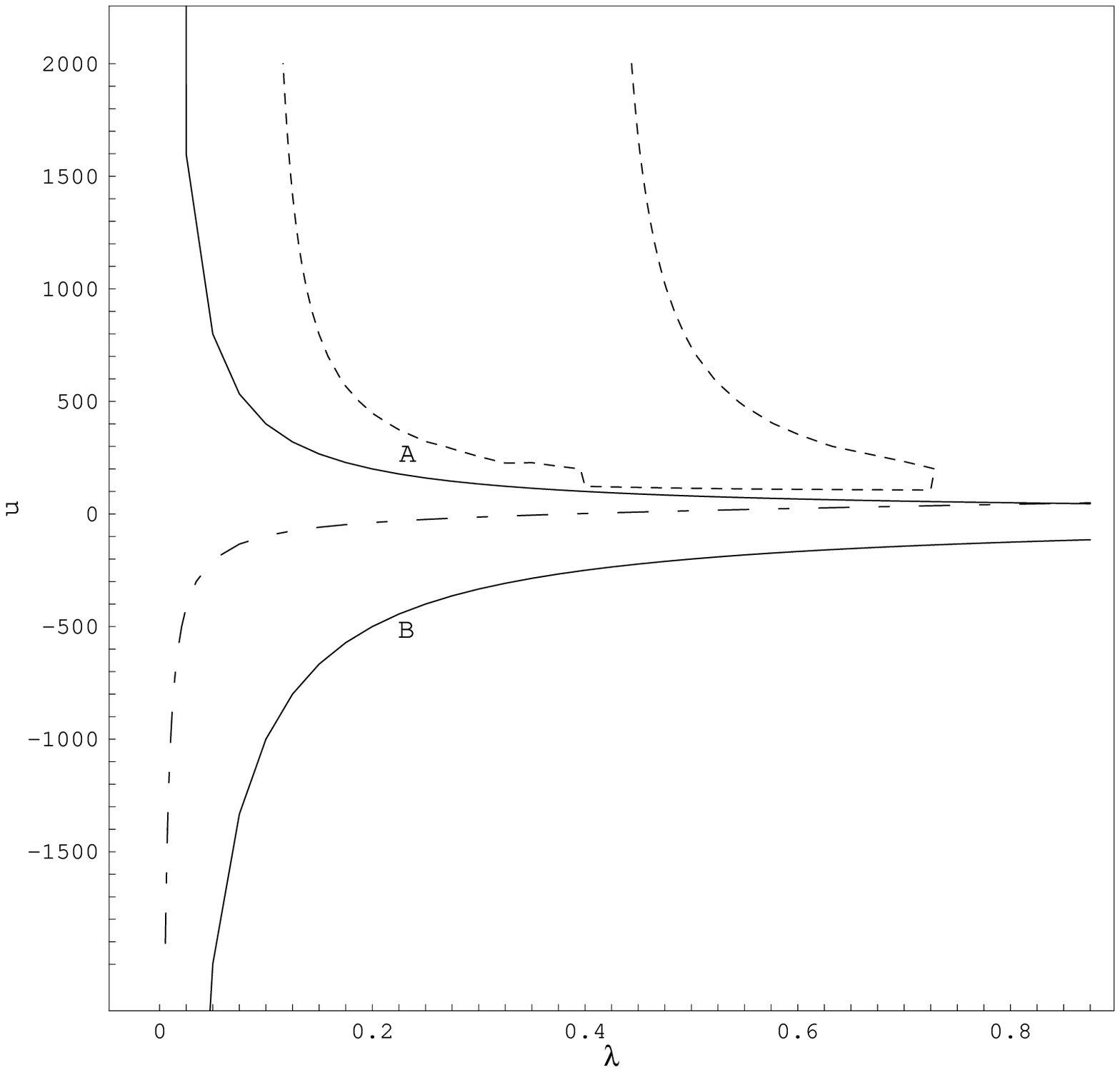}

{\Large\bf Fig. 2}
\end{center}
\end{figure}

\newpage
\setlength{\unitlength}{1mm}
\begin{figure}
\begin{picture}(72,50)(0,0)
\thicklines
\put(35,3){\bf (a)}
\put(58,44){$\tilde\chi_i^0$}
\put(12,44){$\tilde\chi_j^0$}
\put(64,9){$\bar f$}
\put(64,22){$f$}
\put(28,26){$Z_\mu$}
\put(8,40){\vector(1,0){16}}
\put(24,40){\line(1,0){12}}
\put(36,40){\vector(1,0){14}}
\put(50,40){\line(1,0){12}}
\put(36,16){\line(4,-1){13}}
\put(62,9.5){\vector(-4,1){13}}
\put(36,16){\vector(4,1){15}}
\put(62,22.5){\line(-4,-1){11}}
\multiput(36,39)(0,-4){6}{\oval(3,2)[r]}
\multiput(36,37)(0,-4){6}{\oval(3,2)[l]}
\end{picture}\hspace{0.7cm}
\begin{picture}(72,50)(0,0)
\thicklines
\put(35,3){\bf (b)}
\put(58,44){$\bar f$}
\put(12,44){$\tilde\chi_j^0$}
\put(64,9){$\tilde\chi_i^0$}
\put(64,22){$f$}
\put(30,26){$\tilde f$}
\put(8,40){\vector(1,0){16}}
\put(24,40){\line(1,0){12}}
\put(36,40){\line(1,0){14}}
\put(62,40){\vector(-1,0){14}}
\multiput(36,16)(0,5){5}{\line(0,1){3}}
\put(36,16){\vector(4,-1){15}}
\put(62,9.5){\line(-4,1){13}}
\put(36,16){\vector(4,1){15}}
\put(62,22.5){\line(-4,-1){13}}
\end{picture}
\vspace*{1cm}
\begin{center}
{\Large\bf Fig. 3}
\end{center}
\end{figure}

\newpage
\begin{figure}
\setlength{\unitlength}{1mm}
\noindent
\begin{picture}(72,50)(0,0)
\thicklines
\put(35,3){\bf (1a)}
\put(63,44){${\cal A}_\mu$}
\put(12,44){$\tilde\chi_j^0$}
\put(63,10){$\tilde\chi_i^0$}
\put(24,35){${\cal G}_j^{t_\alpha}$}
\put(50,11){${\cal G}_i^{t_\alpha^\ast}$}
\put(39,44){$\tilde t_\alpha$}
\put(56,26){$\tilde t_\alpha^\ast$}
\put(30,29){$t_L$}
\put(41,18){$t_R^c$}
\put(8,40){\vector(1,0){10}}
\put(18,40){\line(1,0){10}}
\multiput(28,40)(5,0){5}{\line(1,0){3}}
\put(38,27){$\bigoplus$}
\put(40,28){\vector(-1,1){6}}
\put(34,34){\line(-1,1){6}}
\put(40,28){\vector(1,-1){6}}
\put(46,22){\line(1,-1){6}}
\multiput(52,16)(0,5){5}{\line(0,1){3}}
\multiput(53,40)(4,0){5}{\oval(2,3)[t]}
\multiput(55,40)(4,0){5}{\oval(2,3)[b]}
\put(62,16){\line(-1,0){10}}
\put(72,16){\vector(-1,0){10}}
\end{picture}\hspace{1cm}
\begin{picture}(72,50)(0,0)
\thicklines
\put(35,3){\bf (1b)}
\put(63,10){${\cal A}_\mu$}
\put(12,44){$\tilde\chi_j^0$}
\put(63,44){$\tilde\chi_i^0$}
\put(24,35){${\cal G}_j^{t_\alpha}$}
\put(52,42){${\cal G}_i^{t_\alpha^\ast}$}
\put(33,44){$\tilde t_\alpha$}
\put(44,44){$\tilde t^\ast_\alpha$}
\put(56,26){$t_R^c$}
\put(34,25){$t_L$}
\put(8,40){\vector(1,0){10}}
\put(18,40){\line(1,0){10}}
\multiput(28,40)(5,0){5}{\line(1,0){3}}
\put(28,40){\line(1,-1){10}}
\put(52,16){\vector(-1,1){14}}
\put(43,22){$\bigoplus$}
\put(52,40){\line(0,-1){12}}
\put(52,16){\vector(0,1){14}}
\put(72,40){\vector(-1,0){10}}
\put(62,40){\line(-1,0){10}}
\multiput(53,16)(4,0){5}{\oval(2,3)[t]}
\multiput(55,16)(4,0){5}{\oval(2,3)[b]}
\end{picture}\vspace{.5cm}\\
\begin{picture}(72,50)(0,0)
\thicklines
\put(35,3){\bf (2a)}
\put(63,44){${\cal A}_\mu$}
\put(12,44){$\tilde\chi_j^0$}
\put(63,10){$\tilde\chi_i^0$}
\put(22,34){${\cal G}_j^{H^-}$}
\put(50,11){${\cal G}_i^{H^+}$}
\put(39,44){$H^-$}
\put(56,26){$H^+$}
\put(28,28){$\tilde\chi^+$}
\put(39,18){$\tilde\chi^-$}
\put(39,26){$\bigoplus$}
\put(8,40){\vector(1,0){10}}
\put(18,40){\line(1,0){10}}
\multiput(28,40)(5,0){5}{\line(1,0){3}}
\put(28,40){\line(1,-1){6}}
\put(40,28){\vector(-1,1){6}}
\put(40,28){\vector(1,-1){6}}
\put(52,16){\line(-1,1){6}}
\multiput(52,16)(0,5){5}{\line(0,1){3}}
\multiput(53,40)(4,0){5}{\oval(2,3)[t]}
\multiput(55,40)(4,0){5}{\oval(2,3)[b]}
\put(52,16){\line(1,0){10}}
\put(72,16){\vector(-1,0){10}}
\end{picture}\hspace{1cm}
\begin{picture}(72,50)(0,0)
\thicklines
\put(35,3){\bf (2b)}
\put(63,10){${\cal A}_\mu$}
\put(12,44){$\tilde\chi_j^0$}
\put(63,44){$\tilde\chi_i^0$}
\put(33,44){$H^-$}
\put(44,44){$H^+$}
\put(56,26){$\tilde\chi^-$}
\put(30,28){$\tilde\chi^+$}
\put(22,34){${\cal G}_j^{H^-}$}
\put(52,42){${\cal G}_i^{H^+}$}
\put(8,40){\vector(1,0){10}}
\put(18,40){\line(1,0){10}}
\multiput(28,40)(5,0){5}{\line(1,0){3}}
\put(28,40){\line(1,-1){12}}
\put(52,16){\vector(-1,1){14}}
\put(52,28){\line(0,1){12}}
\put(52,16){\vector(0,1){14}}
\put(52,40){\line(1,0){10}}
\put(72,40){\vector(-1,0){10}}
\put(43,22){$\bigoplus$}
\multiput(53,16)(4,0){5}{\oval(2,3)[t]}
\multiput(55,16)(4,0){5}{\oval(2,3)[b]}
\end{picture}\vspace{.5cm}\\
\begin{picture}(72,50)(0,0)
\thicklines
\put(35,3){\bf (4a)}
\put(63,44){${\cal A}_\mu$}
\put(12,44){$\tilde\chi_j^0$}
\put(63,10){$\tilde\chi_i^0$}
\put(21,35){${\cal G}_j^{W^-}$}
\put(50,11){${\cal G}_i^{W^+}$}
\put(39,44){$W^-$}
\put(56,26){$W^+$}
\put(30,29){$\tilde\chi^-$}
\put(40,18){$\tilde\chi^+$}
\put(8,40){\vector(1,0){10}}
\put(18,40){\line(1,0){10}}
\multiput(29,40)(4,0){10}{\oval(2,3)[t]}
\multiput(31,40)(4,0){10}{\oval(2,3)[b]}
\put(28,40){\vector(1,-1){7}}
\put(34,34){\line(1,-1){12}}
\put(52,16){\vector(-1,1){6}}
\put(39,26){$\bigoplus$}
\multiput(52,39)(0,-4){6}{\oval(3,2)[r]}
\multiput(52,37)(0,-4){6}{\oval(3,2)[l]}
\put(52,16){\line(1,0){10}}
\put(72,16){\vector(-1,0){10}}
\end{picture}\hspace{1cm}
\begin{picture}(72,50)(0,0)
\thicklines
\put(35,3){\bf (4b)}
\put(63,10){${\cal A}_\mu$}
\put(12,44){$\tilde\chi_j^0$}
\put(63,44){$\tilde\chi_i^0$}
\put(20,35){${\cal G}_j^{W^-}$}
\put(52,42){${\cal G}_i^{W^+}$}
\put(32,44){$W^-$}
\put(43,44){$W^+$}
\put(56,26){$\tilde\chi^+$}
\put(32,26){$\tilde\chi^-$}
\put(8,40){\vector(1,0){10}}
\put(18,40){\line(1,0){10}}
\multiput(29,40)(4,0){6}{\oval(2,3)[t]}
\multiput(31,40)(4,0){6}{\oval(2,3)[b]}
\put(43,22){$\bigoplus$}
\put(28,40){\vector(1,-1){12}}
\put(52,16){\line(-1,1){12}}
\put(52,40){\vector(0,-1){14}}
\put(52,16){\line(0,1){12}}
\put(52,40){\line(1,0){10}}
\put(72,40){\vector(-1,0){10}}
\multiput(53,16)(4,0){5}{\oval(2,3)[t]}
\multiput(55,16)(4,0){5}{\oval(2,3)[b]}
\end{picture}\vspace{.5cm}\\
\begin{picture}(72,50)(0,0)
\thicklines
\put(35,3){\bf (5a)}
\put(63,44){${\cal A}_\mu$}
\put(12,44){$\tilde\chi_j^0$}
\put(63,10){$\tilde\chi_i^0$}
\put(20,35){${\cal G}_j^{G^-}$}
\put(50,11){${\cal G}_i^{W^+}$}
\put(39,44){$G^-$}
\put(56,26){$W^+$}
\put(34,23){$\tilde\chi^+$}
\put(8,40){\vector(1,0){10}}
\put(18,40){\line(1,0){10}}
\multiput(28,40)(5,0){5}{\line(1,0){3}}
\put(52,40){\line(1,0){20}}
\put(28,40){\line(1,-1){12}}
\put(52,16){\vector(-1,1){14}}
\multiput(52,39)(0,-4){6}{\oval(3,2)[r]}
\multiput(52,37)(0,-4){6}{\oval(3,2)[l]}
\put(52,16){\line(1,0){10}}
\put(72,16){\vector(-1,0){10}}
\end{picture}\hspace{1cm}
\begin{picture}(72,50)(0,0)
\thicklines
\put(35,3){\bf (5b)}
\put(63,44){${\cal A}_\mu$}
\put(12,44){$\tilde\chi_j^0$}
\put(63,10){$\tilde\chi_i^0$}
\put(20,35){${\cal G}_j^{W^-}$}
\put(50,11){${\cal G}_i^{G^+}$}
\put(40,44){$W^-$}
\put(56,26){$G^+$}
\put(34,23){$\tilde\chi^-$}
\put(8,40){\vector(1,0){10}}
\put(18,40){\line(1,0){10}}
\multiput(29,40)(4,0){6}{\oval(2,3)[t]}
\multiput(31,40)(4,0){6}{\oval(2,3)[b]}
\put(28,40){\vector(1,-1){14}}
\put(52,16){\line(-1,1){10}}
\multiput(52,16)(0,5){5}{\line(0,1){3}}
\put(52,16){\line(1,0){10}}
\put(72,16){\vector(-1,0){10}}
\multiput(53,40)(4,0){5}{\oval(2,3)[t]}
\multiput(55,40)(4,0){5}{\oval(2,3)[b]}
\end{picture}
\vspace*{.5cm}
\begin{center}
{\Large\bf Fig. 4}
\end{center}
\end{figure}

\newpage
\begin{figure}
\begin{center}
\epsfxsize=\textwidth
\epsfbox{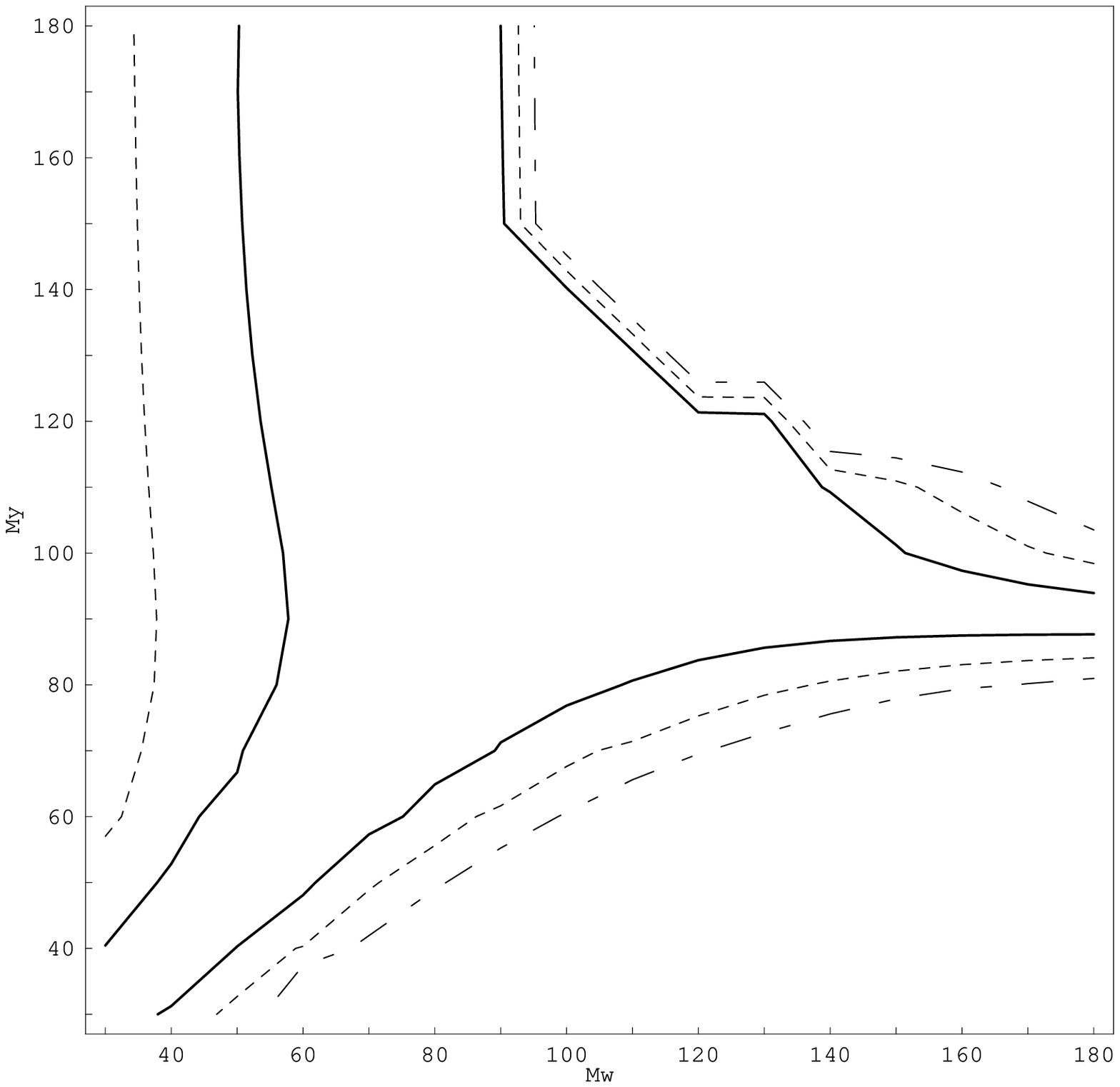}

{\Large\bf Fig. 5a}
\end{center}
\end{figure}

\newpage
\begin{figure}
\begin{center}
\epsfxsize=\textwidth
\epsfbox{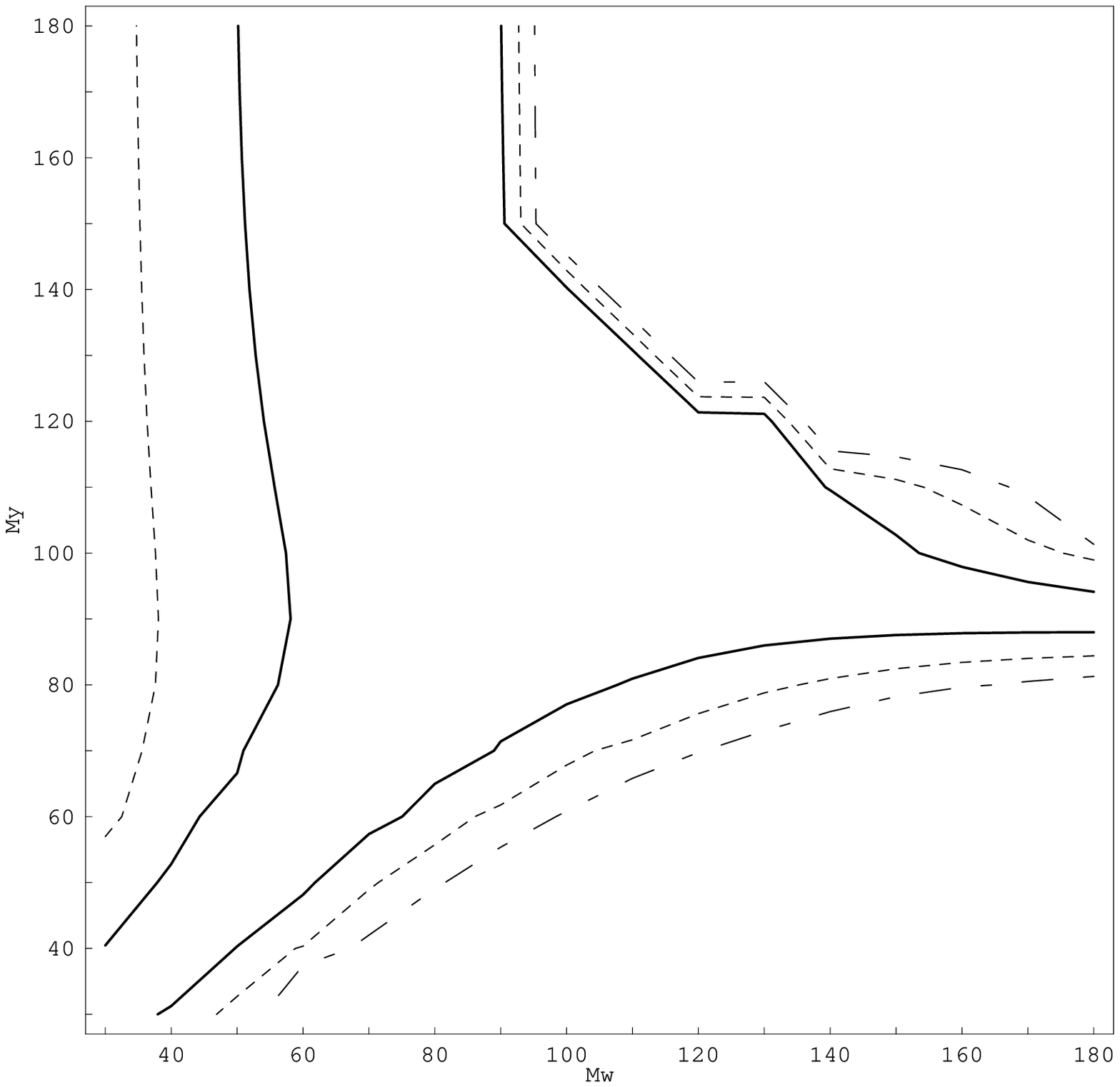}

{\Large\bf Fig. 5b}
\end{center}
\end{figure}

\newpage
\begin{figure}
\begin{center}
\epsfxsize=\textwidth
\epsfbox{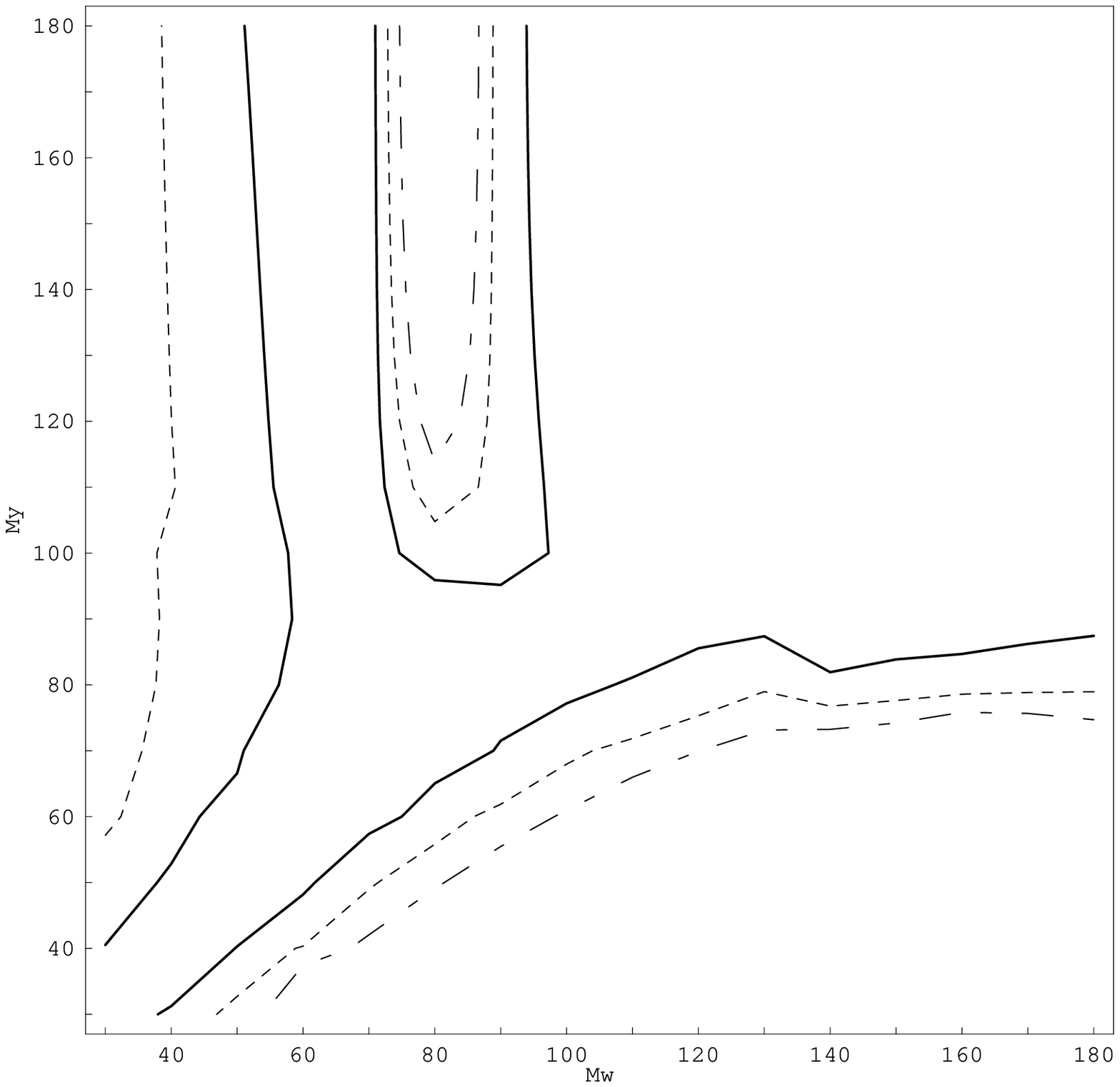}

{\Large\bf Fig. 5c}
\end{center}
\end{figure}

\newpage
\begin{figure}
\begin{center}
\epsfxsize=\textwidth
\epsfbox{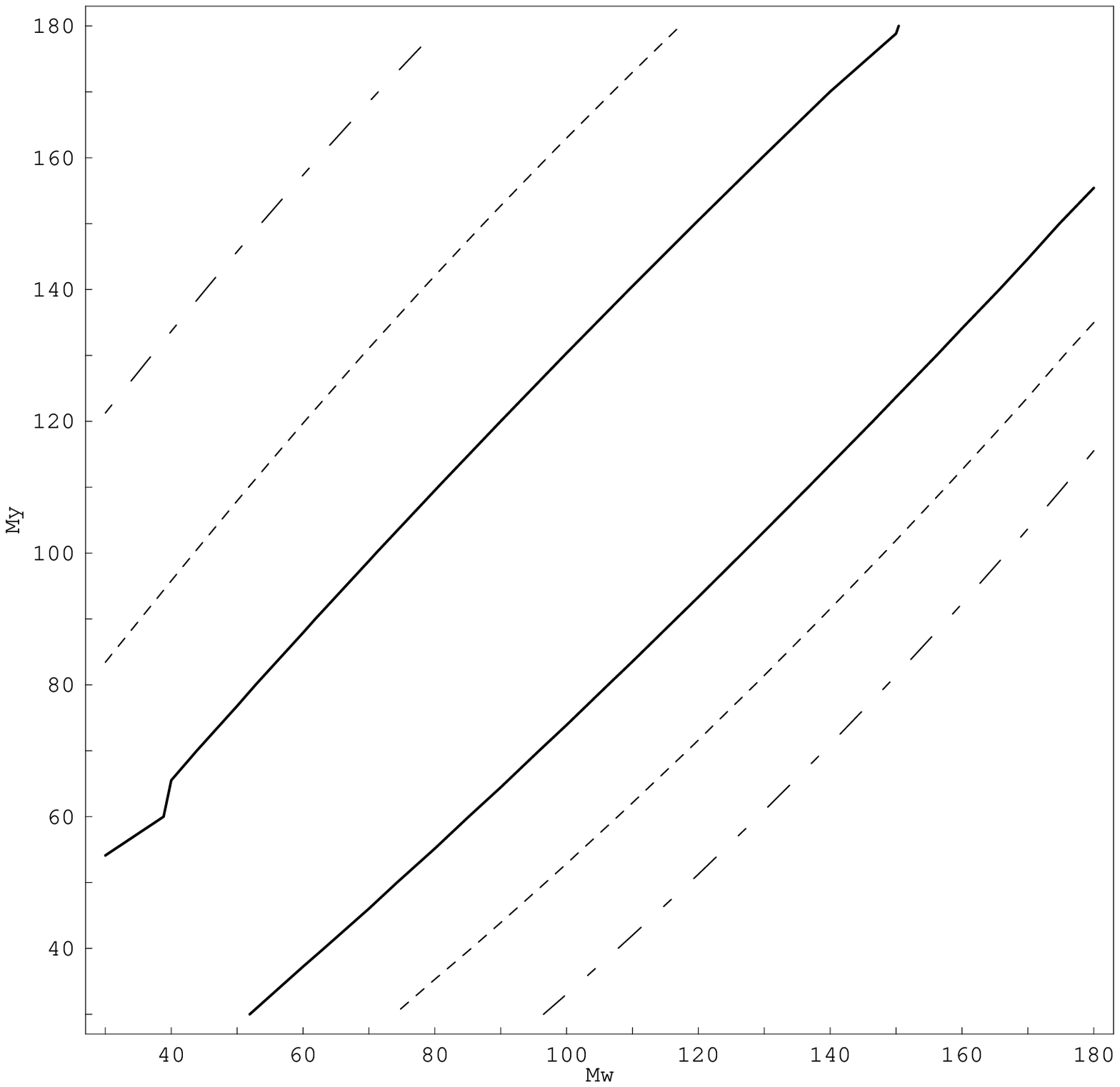}

{\Large\bf Fig. 5d}
\end{center}
\end{figure}

\newpage
\begin{figure}
\begin{center}
\epsfxsize=\textwidth
\epsfbox{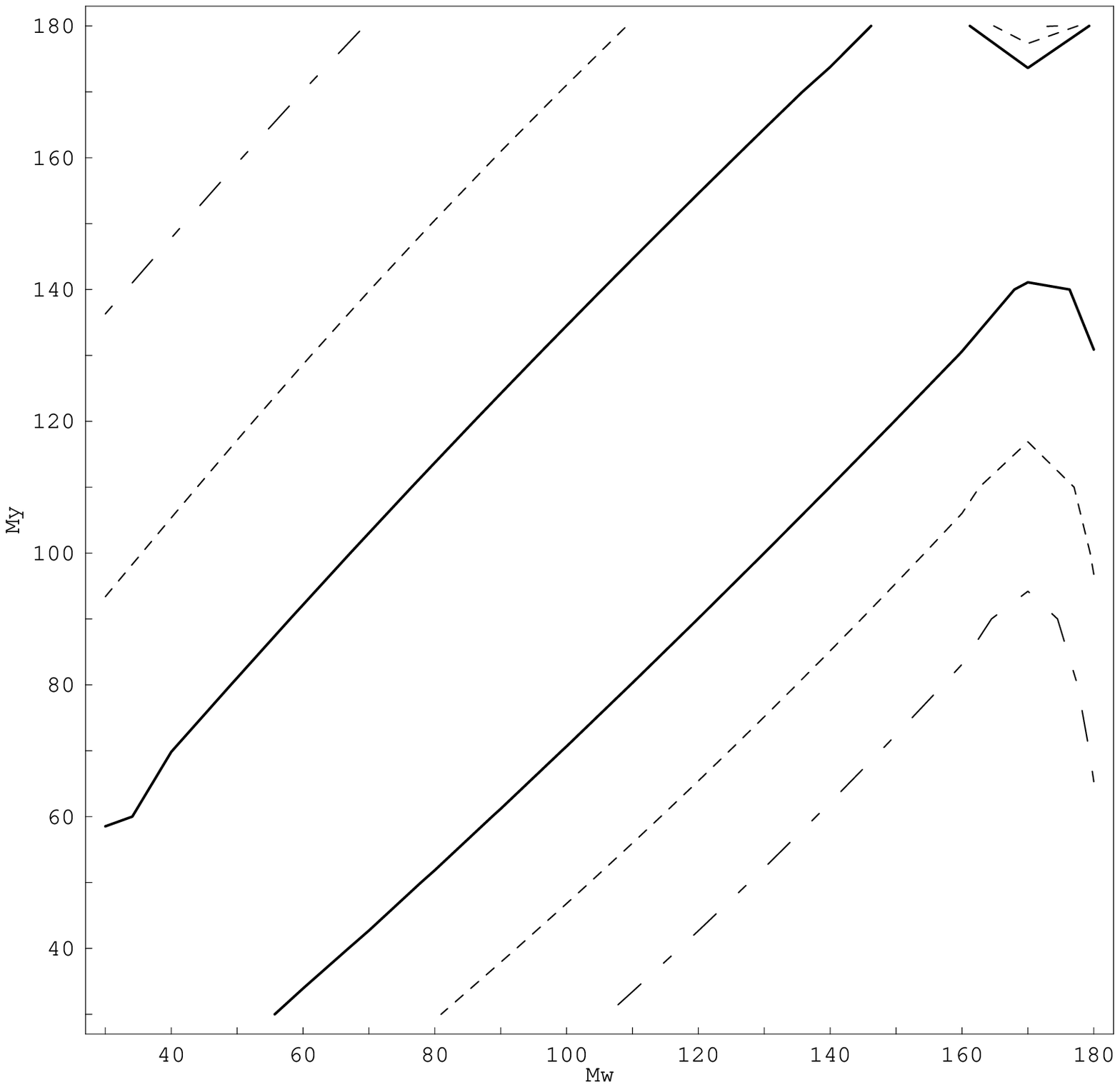}

{\Large\bf Fig. 5e}
\end{center}
\end{figure}

\newpage
\begin{figure}
\begin{center}
\epsfxsize=\textwidth
\epsfbox{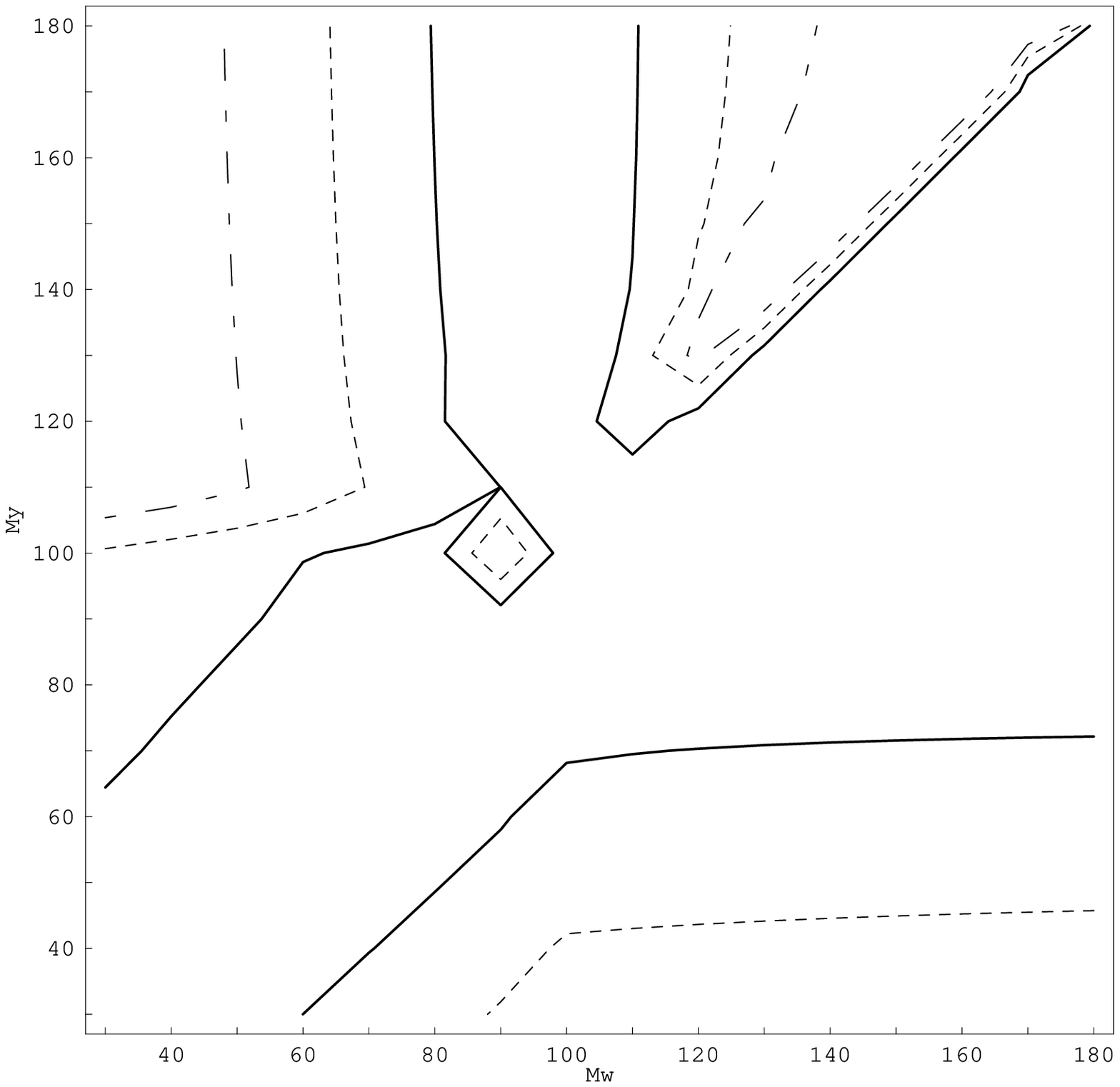}

{\Large\bf Fig. 5f}
\end{center}
\end{figure}


\newpage
\begin{thebibliography}{99}
\bibitem{n}For a review, see for example, \\
H.~-P.~Nilles, Phys. Rep. {\bf C110}, 1 (1984).

H.~E.~Haber and G.~C.~Kane, Phys. Rep. {\bf 117}, 75 (1985).

\bibitem{mu}J.~E.~Kim and H.~P.~Nilles, Phys. Lett. {\bf B138}, 150 (1984).

\bibitem{rad}K.~Inoue, K.~Kakuto, H.~komatsu ans S.~Takeshita,
Prog. theor. Phys. {\bf 68}, 927 (1982).

L.~Alvarez-Gaume, J.~Polchinski and M.~B.~Wise, Nucl. Phys. {\bf B221}, 
495 (1983).

\bibitem{musol}J.~E.~Kim and H.~P.~Nilles, Phys. Lett. {\bf B263},
79 (1991).

E.~J.~Chun, J.~E.~Kim and H.~P.~Nilles, Nucl. Phys. {\bf B370}, 105 (1992).

J.~A.~Casas and C.~Mu\~noz, Phys. Lett {\bf B306}, 228 (1993).

G.~F.~Giudice and A.~Masiero, Phys. Lett. {\bf B206}, 480 (1988).

\bibitem{singlet}H.~P.~Nilles, M.~Srednicki and D.~Wyler, Phys. Lett.
{\bf B129}, 364 (1983).

J.-P.~Derendinger and C.~A.~Savoy, Nucl. Phys. {\bf B237}, 364 (1984).

L.~E.~Ib\'a\~nez and J.~Mas, Nucl. Phys. {\bf B286}, 107 (1987).

J.~Ellis, J.~F.~Gunion, H.~E.~Haber, L.~Roszkowski and F.~Zwirner,
Phys. Rev. {\bf D39}, 844 (1989).

\bibitem{sy}D.~Suematsu and Y.~Yamagishi, Int. J. Mod. Phys. {\bf A10}, 
4521 (1995).

\bibitem{grh}J.~F.~Gunion, L.~Roszkowski and H.~E.~Haber,
Phys. Rev. {\bf D38}, 105 (1988).

\bibitem{extra}For recent works, see for example,
M.~Cveti$\check{\rm c}$ and P.~Langacker, Phys. Rev. 
{\bf D54}, 3570 (1996).

T.~Gherghetta, T.~A.~Kaeding and G.~L.~Kane, preprint UM-TH-96-18
(hep-ph/9701343).

M.~Cveti$\check{\rm c}$, D.~A.~Demir, J.~R.~Espinosa, L.~Everett and 
P.~Langacker, preprint UPR-0737-T (hep-ph/973317).

\bibitem{string}For a review, see for example, \\
J.~Hewett and T.~G.~Rizzo, Phys. Rep. {\bf C183}, 193 (1989).

F.~Zwirner, Int. J. Mod. Phys. {\bf A3}, 49 (1988).

\bibitem{exp}For example,\\
M.~Cveti$\check{\rm c}$ and S.~Godfrey, in Proceedings of Electro-weak Symmetry
Breaking and beyond the standard model, eds T.~Barklow, S.~Dawson,
H.~Haber and J.~Siegrist (World Scientific 1995), hep-ph/9504216.
 
K.~Maeshima, Proceedings of the 28th International Conference on High
Energy Physics (ICHEP'96), Warsaw, Poland, 1996.

\bibitem{sue}D.~Suematsu, preprint KANAZAWA-97-06 (hep-ph/9705405).

D.~Suematsu, Mod. Phys. Lett. {\bf A12}, 1709 (1997), (hep-ph/97055412).

\bibitem{mssmdecay}H.~Komatsu and J.~Kubo, Phys. Lett. {\bf 157B}, 90(1985);
Nucl. Phys. {\bf B263}, 265 (1986).

H.~E.~Haber and D.~Wyler, Nucl. Phys. {\bf B323}, 267 (1989).

\bibitem{cdf}S.~Ambrosanio, G.~L.~Kane, G.~D.~Kribs, 
S.~P.~Martin and S.~Mrenna, Phys. Rev. Lett. {\bf 76}, 3498 (1996);
Phys. Rev.{\bf D55}, 1372 (1997).

\bibitem{am}S.~Ambrosanio and B.~Mele, Phys. Rev. {\bf D55}, 1399 (1997).

\bibitem{mixing}B.~Holdom, Phys. Lett. {\bf B166}, 196 (1986).

K.~Choi and J.~E.~Kim, Phys. Lett. {\bf 165B}, 71 (1985).

\bibitem{ms}T.~Matsuoka and D.~Suematsu, Prog. Theor. Phys. {\bf 76},
 901 (1986).

\bibitem{mixing2}K.~S.~Babu, C.~Kolda and J.~March-Russell,
Phys. Rev. {\bf D54}, 4635 (1996).

K.~R.~Dienes, C.~Kolda and J.~March-Russell, preprint IASSNS-HEP-96/100
(hep-ph/9610479).

\bibitem{discr}N.~Haba, C.~Hattori, M.~Matsuda, T.~Matsuoka and
D.~Mochinaga, Prog. Theor. Phys. {\bf 92}, 153 (1994).

G,~Cleaver, M.Cveti$\check{\rm c}$, J.~R.~Espinosa, L.~Everett and
P.~Langacker, preprint UPR-0750-T(hep-ph/9705391).

\bibitem{dflat}M.~Dine, V.~Kaplunovsky, M.~Mangano, C.~Nappi and
N.~Seiberg, Nucl. Phys. {\bf B254}, 549 (1985).

T.~Matsuoka and D.~Suematsu, Nucl. Phys. {\bf B274},
106 (1986); Prog. Theor. Phys. {\bf 76}, 886 (1986).

\bibitem{sue0}D.~Suematsu, Prog. Theor. Phys. {\bf 96}, 611 (1996).

\bibitem{neut}D.~Suematsu, Phys. Lett. {\bf B392}, 413 (1997)
~(hep-ph/9604242). 

\bibitem{infl}D.~Suematsu and Y.~Yamagishi, Mod. Phys. Lett. {A10},
2923 (1995).

\bibitem{acq}F.~del Aguila, G.~D.~Coughlan and M.~Quir\'os,
Nucl. Phys. {\bf B307}, 633 (1988).

\bibitem{mixxi}L3 Collaboration, Phys. Lett. {\bf B306}, 187 (1993).

DELPHI Collaboration, Z. Physik {\bf C65}, 603 (1995).

\bibitem{hs}H.~E.~Haber and M.~Sher, Phys. Rev. {\bf D35}, 2206
(1987).

\bibitem{drees}M.~Drees, Phys. Rev. {\bf D35}, 2910 (1987).
 
\bibitem{pdata}Particle Data Group, Phys. Rev. {\bf D54}, 225 (1996).

\bibitem{aleph} OPAL collaboration, Phys. Lett. {\bf B389}, 616 (1996).

ALEPH collaboration, Z. Physik {\bf C72}, 549 (1996).

\bibitem{ndecay}A.~H~.Chamseddine, P.~Nath and R.~Arnowitt,
Phys. Lett. {\bf 129B}, 445 (1983).

J.~Ellis, J.-M.Fr\`ere,J.~S.~Hagelin, G.~L.~Kane and S.~T.~Petcov,
Phys. Lett. {\bf 132B}, 436 (1983).

A.~Bartl, H.~Fraas and W.~Majerotto, Nucl. Phys. {\bf B278}, 1 (1986).
 
\bibitem{effect}N.~Cabibbo, G.~R.~Farrar and L.~Maiani, Phys. Lett. 
{\bf 105B}, 155 (1981).

\bibitem{ce}B.~de Carlos and J.~R.~Espinosa, preprint SUSX-TH-97-008 
(hep-ph/9705315).

\end{thebibliography}
\end{document}